\documentclass[11pt]{article}

\usepackage{acl}
\usepackage{comment} % Required package
\usepackage{times}
\usepackage{latexsym}

\usepackage[T1]{fontenc}
\usepackage[utf8]{inputenc}

\usepackage{microtype}

\usepackage{inconsolata}

\usepackage{graphicx}
\usepackage{multirow}
\usepackage{booktabs}
\usepackage{array}
\usepackage{float}
\usepackage{graphicx} % for \rotatebox
\usepackage{adjustbox}
\usepackage{multirow}
\usepackage{booktabs}
\usepackage{array} % 提供 m{..} 与 \arraybackslash
\newcolumntype{M}[1]{>{\centering\arraybackslash}m{#1}}
\usepackage{amsmath}
\usepackage{amsfonts}
\usepackage{graphicx}

\title{Adaptive Hierarchical Representation Alliance for Multimodal Learning}

\author{
 \textbf{Chunlei Meng\textsuperscript{1}},
 \textbf{Pengbin Feng\textsuperscript{2}},
 \textbf{Jacqueline J. Pang\textsuperscript{3}}\\
 \textbf{Chih-Ting Liao\textsuperscript{4}},
 \textbf{Rong Fu\textsuperscript{5}},
 \textbf{Zhaolu Kang\textsuperscript{6}},
 \textbf{Zhongxue Gan\textsuperscript{1}},
 \textbf{Chun Ouyang*\textsuperscript{1}}
\\
 \textsuperscript{1}College of Intelligent Robotics and Advanced Manufacturing, Fudan University\\
 \textsuperscript{2}University of Southern California,
 \textsuperscript{3}Cornell University \\
 \textsuperscript{4}University of New South Wales,
 \textsuperscript{5}Independent Researcher,
 \textsuperscript{6}Peking University
\\
\small{
\textbf{*Correspondence:}
\href{mailto:oy_c@fudan.edu.cn}{oy\_c@fudan.edu.cn}.
\href{mailto:clmeng23@m.fudan.edu.cn}{clmeng23@m.fudan.edu.cn}
}\\
\small{
\textbf{This study has been accepted by EMNLP 2026 (Findings)}}
 }

\begin{document}
\maketitle

\begin{abstract}
Multimodal models often align language, vision, and audio in a single final-layer latent space, implicitly assuming that task-relevant evidence emerges at the same semantic depth across modalities. Using layer-wise CKA analysis, we observe that this assumption leads to semantic granularity mismatch: textual cues usually require deeper contextual abstraction, whereas visual and acoustic cues often provide discriminative perceptual evidence in shallow or middle layers. This mismatch can flatten fine-grained modality-private cues and reduce reliability under noisy, imbalanced, or missing inputs. To address this, we proposed Adaptive Hierarchical Representation Alliance (AHRA), a hierarchical shared--private expert framework. AHRA factorizes each modality into shared and private streams across semantic levels, regularizes them with shared alignment and private decorrelation, routes shared information through a cross-modal expert, and enhances task-relevant private tokens with modality-specific experts guided by a sparsity-controlled
soft-gating mechanism (foreground exam). A hierarchical co-fusion module then performs intra-level expert coordination and inter-level semantic selection. Experiments on six benchmarks across image-text classification, multimodal intent recognition, and trimodal sentiment analysis show that AHRA consistently improves over strong baselines and remains robust under noisy and missing-modality settings.
\end{abstract}

\section{Introduction}
\label{Intro}
Multimodal Learning (MML) integrates heterogeneous signals such as language, vision, and audio into unified representations for tasks including sentiment analysis, intent recognition, and multimodal classification~\cite{TSD,TSDA}. The central challenge is modality heterogeneity: different signals vary in temporal scale, statistical distribution, and abstraction level~\cite{EMOE,I2RL,DBR}. Existing methods mainly address this challenge through two directions. Representation disentanglement separates modality-invariant and modality-specific factors~\cite{misa,DLF,CLCR,MRCF,confede,MGJR}, while dynamic or robust fusion calibrates modality contributions under noise, imbalance, or missing inputs~\cite{MLA,QMF,EAU,DEVA,I2RL,PEFT,GCL}. Although effective, most methods still perform alignment, calibration, and prediction in a single terminal latent space. This flat protocol implicitly assumes that task-relevant evidence from all modalities emerges at the same semantic depth.

\begin{figure}[t]
\centering
\includegraphics[width=\columnwidth]{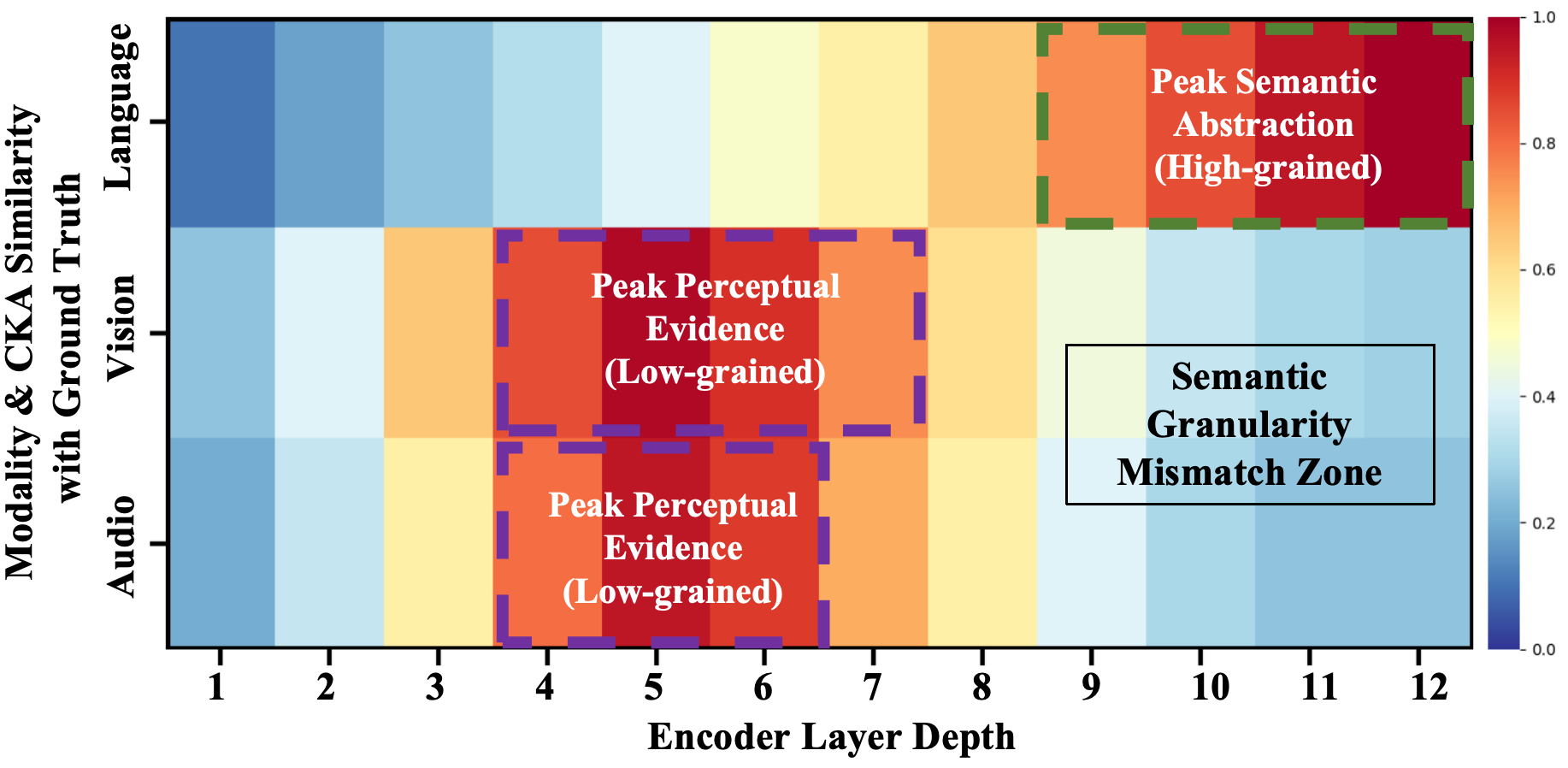}
\caption{Layer-wise CKA similarity with labels. Text peaks in deep layers, while vision and audio peak in shallow or middle layers, revealing Semantic Granularity Mismatch (SGM) and motivating hierarchical fusion.}
\label{fig:SGM}
\end{figure}

We argue that this assumption overlooks \textbf{Semantic Granularity Mismatch (SGM)}. Language is symbolic and compositional, so its discriminative semantics often require deep contextual abstraction. In contrast, visual and acoustic signals are perceptual: cues such as facial movements, objects, tone, or prosody may already be informative in shallow or middle layers. To examine this divergence, we conduct a layer-wise semantic analysis using Centered Kernel Alignment (CKA)~\cite{CKA} between hidden representations and task labels:
\begin{equation}
\mathrm{CKA}(K,L)
=
\frac{
\mathrm{HSIC}(K,L)
}{
\sqrt{
\mathrm{HSIC}(K,K)\mathrm{HSIC}(L,L)
}
},
\label{eq:cka_intro}
\end{equation}
where $K$ and $L$ denote kernel matrices constructed from layer-wise representations and labels, respectively, and HSIC denotes the Hilbert--Schmidt Independence Criterion. As shown in Fig.~\ref{fig:SGM}, text reaches its strongest label correlation in deep layers, whereas vision and audio peak earlier. Final-layer alignment can force low-granularity perceptual evidence into an overly abstract space, causing useful modality-private details to be diluted.

SGM leads to two practical failures. First, feature flattening suppresses fine-grained perceptual cues as background variation when they are projected into deep semantic spaces. Second, inference rigidity makes the model less able to rely on trustworthy unimodal evidence when modalities conflict, are corrupted, or are partially missing. These failures suggest that multimodal fusion should not only align modalities globally, but preserve and select evidence across different semantic levels.

\begin{figure*}[t]
\centering
\includegraphics[width=\linewidth]{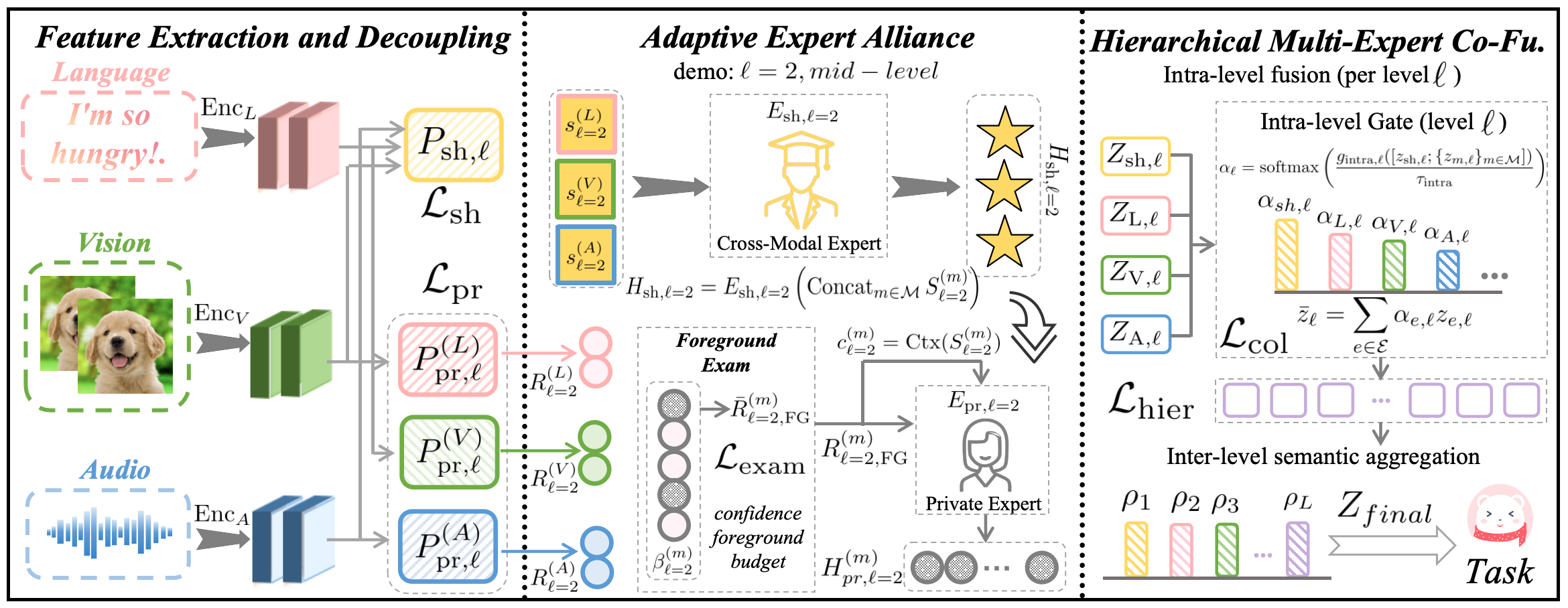}
\caption{Overview of AHRA. The trimodal case is shown for illustration. For each modality $m\in\mathcal{M}$ and semantic level $\ell$, AHRA factorizes encoder features into shared streams $S_\ell^{(m)}$ and private streams $R_\ell^{(m)}$, routes them through a shared-space cross-modal expert and foreground-controlled modality-specific experts, and obtains $z_{\mathrm{final}}$ via intra-level expert co-fusion and inter-level semantic aggregation.}
\label{fig:AHRA}
\end{figure*}

To address this problem, we propose \textbf{Adaptive Hierarchical Representation Alliance (AHRA)}. As illustrated in Fig.~\ref{fig:AHRA}, AHRA extracts multi-level features from each modality and factorizes them into shared and private streams at every semantic level. The shared streams capture cross-modal agreement, while the private streams retain modality-specific cues that may be lost under single-space fusion. This decomposition is regularized by shared alignment and private decorrelation, preventing the model from collapsing back into a single latent space. On top of these streams, an Adaptive Expert Alliance (AEA) applies a cross-modal expert to shared representations and modality-specific experts to private representations, with a sparsity-regularized foreground exam softly emphasizing task-relevant private tokens. A Hierarchical Multi-Expert Co-Fusion module (HMEC) then performs intra-level expert coordination and inter-level semantic selection to produce the final multimodal representation. More related works see Appendix~\ref{app-section:Related work}. Our contributions are as follows:

% \caption{Overview of AHRA. The model factorizes multi-level multimodal features into shared and private streams, enhances them with shared and modality-specific experts, and aggregates evidence through hierarchical co-fusion.}

\begin{itemize}
    \item We identify \textbf{Semantic Granularity Mismatch (SGM)} as a structural limitation of single-space multimodal fusion, and provide layer-wise CKA evidence showing that language, vision, and audio reach peak label correlation at different semantic depths.

    \item We propose \textbf{Adaptive Hierarchical Representation Alliance (AHRA)}, a hierarchical shared-private expert framework that preserves modality-private cues across multiple semantic levels while maintaining cross-modal agreement.

    \item We design \textbf{Adaptive Expert Alliance (AEA)} and \textbf{Hierarchical Multi-Expert Co-Fusion (HMEC)} to perform adaptive private-token enhancement, intra-level expert coordination, and inter-level semantic selection, leading to strong performance on six multimodal benchmarks across image-text classification, multimodal intent recognition, and trimodal sentiment analysis, including noisy and missing-modality settings.
\end{itemize}

\section{Methodology}
\label{section:Methodology}
\subsection{Model Overview}

We consider supervised multimodal learning over a set of dataset-level
modalities $\mathcal{M}\subseteq\{L,V,A\}$, where $L$, $V$, and $A$ denote
language, vision, and audio. For trimodal datasets, $\mathcal{M}=\{L,V,A\}$;
for image--text datasets, $\mathcal{M}=\{L,V\}$. Let
$(m_1,\ldots,m_{|\mathcal{M}|})$ be a fixed modality ordering and define
\begin{equation}
\mathcal{P}=\{(m_a,m_b):1\le a<b\le |\mathcal{M}|\}.
\end{equation}
Here $\mathcal{M}$ refers to dataset-level modalities; sample-level missing or
imbalanced modalities are evaluated in the robustness experiments.

Adaptive Hierarchical Representation Alliance (AHRA) addresses semantic
granularity mismatch by combining multi-level shared--private factorization with
adaptive expert routing. At each semantic level, AHRA projects each modality
into a shared stream and a private stream. The shared streams are regularized
for cross-modal coordinate compatibility, while the private streams are
regularized only through linear decorrelation rather than a hard independence
constraint. An Adaptive Expert Alliance (AEA) routes shared streams to a
cross-modal expert and foreground-weighted private streams to modality-specific
experts. A Hierarchical Multi-Expert Co-Fusion module (HMEC) then performs
intra-level expert coordination and inter-level semantic aggregation. 

\subsection{Semantic Granularity Diagnostic}

AHRA is motivated by semantic granularity mismatch: different modalities may
reach their strongest task-dependent representations at different encoder
depths. For diagnostic CKA, $\ell$ indexes evaluated encoder depths; for AHRA
training, selected depths are grouped into $L$ semantic levels, e.g., shallow,
middle, and deep. For modality $m$, depth/level $\ell$, and sample $i$, we define
$h^{(m)}_{\ell,i}=\operatorname{Pool}(H^{(m)}_{\ell,i})$,
$D_{m,\ell}=\operatorname{Dep}(\{h^{(m)}_{\ell,i}\}_{i=1}^{n},\{y_i\}_{i=1}^{n})$,
$\ell_m^\star=\arg\max_{\ell}D_{m,\ell}$.
In our diagnostic analysis, $\operatorname{Dep}(\cdot,\cdot)$ is estimated by
Centered Kernel Alignment (CKA) between representation and label kernels. The
semantic granularity mismatch score is
\begin{equation}
\mathrm{SGM} =
\frac{1}{|\mathcal{P}|}
\sum_{(m,m') \in \mathcal{P}}
\left|
\frac{\ell_m^{*}}{L_m}
-
\frac{\ell_{m'}^{*}}{L_{m'}}
\right|.
\end{equation}
The CKA analysis is used as an empirical diagnostic rather than a formal proof;
it motivates retaining multiple semantic levels instead of relying only on
final-layer fusion.

\subsection{Multi-Level Shared--Private Factorization}

\paragraph{Semantic levels and projections.}
For each modality $m\in\mathcal{M}$, a unimodal encoder extracts $L$ semantic
levels,
\begin{equation}
\{H^{(m)}_{\ell}\}_{\ell=1}^{L}
=
\operatorname{Enc}_m(x^{(m)}),
\quad
H^{(m)}_{\ell}\in\mathbb{R}^{T^{(m)}_{\ell}\times d}.
\end{equation}
If raw backbone widths differ, the corresponding width-matching adapter is
absorbed into $\operatorname{Enc}_m(\cdot)$, so all $H^{(m)}_{\ell}$ have the
common width $d$. At level $\ell$, AHRA applies a modality-agnostic shared
projector and modality-specific private projectors:
\begin{equation}
S^{(m)}_{\ell}=H^{(m)}_{\ell}P_{\mathrm{sh},\ell},
\quad
R^{(m)}_{\ell}=H^{(m)}_{\ell}P^{(m)}_{\mathrm{pr},\ell},
\label{eq:shared_private}
\end{equation}
where $P_{\mathrm{sh},\ell}\in\mathbb{R}^{d\times d_{\mathrm{sh}}}$ and
$P^{(m)}_{\mathrm{pr},\ell}\in\mathbb{R}^{d\times d_{\mathrm{pr}}}$. The shared
stream captures cross-modal common factors, while the private stream preserves
modality-specific residual factors. These losses encourage separation but do not
assume statistically identifiable disentanglement.

\paragraph{Shared alignment.}
For each sample, we mean-pool the shared tokens and standardize each coordinate
over the mini-batch:
\begin{equation}
s^{(m)}_{\ell,i}
=
\frac{1}{T^{(m)}_{\ell}}
\sum_{t=1}^{T^{(m)}_{\ell}}S^{(m)}_{\ell,i}[t,:],
\hat{s}^{(m)}_{\ell,i,j}
=
\frac{s^{(m)}_{\ell,i,j}-\mu^{(m)}_{\ell,j}}
{\sigma^{(m)}_{\ell,j}+\epsilon}.
\end{equation}
This is dimension-wise standardization and does not whiten the full covariance
matrix. The empirical cross-modal correlation matrix is
\begin{equation}
C^{(m,m')}_{\mathrm{sh},\ell}
=
\frac{1}{B}
\sum_{i=1}^{B}
\hat{s}^{(m)}_{\ell,i}
\hat{s}^{(m')\top}_{\ell,i}
\in\mathbb{R}^{d_{\mathrm{sh}}\times d_{\mathrm{sh}}}.
\end{equation}
AHRA uses the identity-shaped correlation loss
\begin{equation}
\mathcal{L}_{\mathrm{sh}}
=
\frac{1}{L|\mathcal{P}|}
\sum_{\ell=1}^{L}
\sum_{(m,m')\in\mathcal{P}}
\left\|
C^{(m,m')}_{\mathrm{sh},\ell}-I
\right\|_{F}^{2}.
\end{equation}
The diagonal terms encourage coordinate-wise cross-modal agreement, and the
off-diagonal terms reduce cross-coordinate redundancy. Since
$C^{(m,m')}_{\mathrm{sh},\ell}$ is estimated from finite mini-batches, this loss
is used as a stochastic correlation-shaping regularizer rather than an exactly
attainable hard constraint in every batch.

\paragraph{Private linear decorrelation.}
Applying the same pooling and standardization to private streams gives
$\hat r^{(m)}_{\ell,i}$. The private regularizer contains intra-modality
shared--private decorrelation and cross-modality private--private decorrelation:

{
\small
\begin{align}
\mathcal{L}^{\mathrm{intra}}_{\mathrm{pr}}
&=
\frac{1}{L|\mathcal{M}|}
\sum_{\ell=1}^{L}
\sum_{m\in\mathcal{M}}
\left\|
\frac{1}{B}
\sum_{i=1}^{B}
\hat{s}^{(m)}_{\ell,i}\hat{r}^{(m)\top}_{\ell,i}
\right\|_{F}^{2},\\
\mathcal{L}^{\mathrm{cross}}_{\mathrm{pr}}
&=
\frac{1}{L|\mathcal{P}|}
\sum_{\ell=1}^{L}
\sum_{(m,m')\in\mathcal{P}}
\left\|
\frac{1}{B}
\sum_{i=1}^{B}
\hat{r}^{(m)}_{\ell,i}\hat{r}^{(m')\top}_{\ell,i}
\right\|_{F}^{2}.
\end{align}
}

Then
\begin{equation}
\mathcal{L}_{\mathrm{dec}}
=
\lambda_{\mathrm{sh}}\mathcal{L}_{\mathrm{sh}}
+
\lambda_{\mathrm{pr}}
\left(
\mathcal{L}^{\mathrm{intra}}_{\mathrm{pr}}
+
\mathcal{L}^{\mathrm{cross}}_{\mathrm{pr}}
\right).
\label{eq:loss_dec}
\end{equation}
The private loss penalizes linear cross-correlation and reduces duplicated
linear factors, but it is not claimed to guarantee full statistical
independence. Private streams remain task-grounded through the downstream task
loss and private experts.

\subsection{Adaptive Expert Alliance}

AEA separates cross-modal reasoning from modality-private enhancement. For each
semantic level, it constructs one shared-space cross-modal expert and one
private expert for each available modality.

\paragraph{Shared-space cross-modal expert.}
At level $\ell$, shared tokens are concatenated and passed to a cross-modal
expert:
\begin{equation}
H_{\mathrm{sh},\ell}
=
E_{\mathrm{sh},\ell}
\left(
\operatorname{Concat}_{m\in\mathcal{M}}S^{(m)}_{\ell}
\right).
\end{equation}
The expert is implemented as a lightweight attention block, so cross-modal
interactions are input-dependent but restricted to the shared subspace.

\paragraph{Private-space foreground exam.}
For modality $m$ at level $\ell$, each private token receives a soft foreground
score:
\begin{equation}
\beta^{(m)}_{\ell,i,t}
=
\sigma\left(
(w^{(m)}_{\ell})^\top r^{(m)}_{\ell,i,t}
+
b^{(m)}_{\ell}
\right).
\end{equation}

\begin{equation}
R^{(m)}_{\ell,\mathrm{FG}}
=
\left\{
\beta^{(m)}_{\ell,i,t}
r^{(m)}_{\ell,i,t}
\right\}_{t=1}^{T^{(m)}_{\ell}}.
\end{equation}
A modality-specific expert enhances the foreground-weighted private stream under
same-level shared context:
\begin{equation}
c^{(m)}_{\ell}=\operatorname{Ctx}(S^{(m)}_{\ell}),
H^{(m)}_{\mathrm{pr},\ell}
=
E^{(m)}_{\mathrm{pr},\ell}
\left(
R^{(m)}_{\ell,\mathrm{FG}};\,
c^{(m)}_{\ell}
\right).
\end{equation}
Here $\operatorname{Ctx}(\cdot)$ denotes the shared-context representation used
by the implementation, and the semicolon indicates conditioning rather than a
separate prediction branch. This keeps the private expert anchored in
foreground-weighted private tokens while allowing shared context to calibrate
private-token enhancement.

\paragraph{Foreground budget and confidence.}
To avoid notation conflict with inter-level weights, we denote the target
foreground ratio by $\pi_{m,\ell}$. The average foreground rate is
$\bar{\beta}_{m,\ell}
=
\mathbb{E}_{i,t}
[
\beta^{(m)}_{\ell,i,t}
]$.
The exam regularizer is
\begin{equation}
\begin{aligned}
\mathcal{L}_{\mathrm{exam}}
&=
\sum_{\ell=1}^{L}
\sum_{m\in\mathcal{M}}
\Bigg[
\lambda_{\mathrm{rate}}
(\bar{\beta}_{m,\ell}-\pi_{m,\ell})^{2}
\\
&\quad+
\lambda_{\mathrm{conf}}
\mathbb{E}_{i,t}
\big[
\beta^{(m)}_{\ell,i,t}
(1-\beta^{(m)}_{\ell,i,t})
\big]
\Bigg]
\end{aligned}
\label{eq:loss_exam}
\end{equation}
The rate term controls the soft token budget, and the confidence term
discourages ambiguous gates near $0.5$.

\subsection{Hierarchical Multi-Expert Co-Fusion}

HMEC factorizes routing into intra-level expert coordination and inter-level
semantic aggregation. For compactness, we omit the sample index in the following
routing equations; all expert and level weights are computed per sample.

\paragraph{Intra-level expert co-fusion.}
At level $\ell$, the shared expert output and private expert outputs are pooled
into descriptors $z_{\mathrm{sh},\ell}$ and $\{z_{m,\ell}\}_{m\in\mathcal{M}}$.
Let $\mathcal{E}=\{\mathrm{sh}\}\cup\mathcal{M}$ denote the expert index set. The
intra-level gate is
\begin{equation}
\alpha_{\ell}
=
\operatorname{softmax}
\left(
\frac{
g_{\mathrm{intra},\ell}
([z_{\mathrm{sh},\ell};\{z_{m,\ell}\}_{m\in\mathcal{M}}])
}
{\tau_{\mathrm{intra}}}
\right)
\end{equation}
and $\bar z_{\ell}
=
\sum_{e\in\mathcal{E}}
\alpha_{e,\ell}z_{e,\ell}.$
\paragraph{Inter-level semantic aggregation.}
The inter-level gate assigns semantic-level weights and produces the final
representation:
\begin{equation}
\rho
=
\operatorname{softmax}
\left(
\frac{
\psi_{\mathrm{inter}}([\bar z_1;\ldots;\bar z_L])
}
{\tau_{\mathrm{inter}}}
\right)
\end{equation}
and $z_{\mathrm{final}}
=
\sum_{\ell=1}^{L}
\rho_{\ell}\bar z_{\ell}.$
Equivalently, HMEC factorizes the joint routing distribution as
\[
p(e,\ell\mid x)=p(\ell\mid x)p(e\mid \ell,x),
\]
separating expert selection from semantic-level aggregation.

\paragraph{Gate regularization.}
For intra-level gates, AHRA discourages systematic under-use of private experts:
% \begin{equation}
% \mathcal{L}_{\mathrm{col}}
% =
% \lambda_{\mathrm{fair}}
% \sum_{m\in\mathcal{M}}
% \left(
% \mathbb{E}_{x,\ell}[\alpha_{m,\ell}]
% -
% \eta_m
% \right)^2.
% \end{equation}
\begin{equation}
\mathcal{L}_{\mathrm{col}}
=
\lambda_{\mathrm{fair}}
\sum_{m \in \mathcal{M}}
\left[
\max(0, \eta_m - \mathbb{E}_{x,\ell}[\alpha_{m,\ell}])
\right]^2
\end{equation}
Here $\eta_m$ is a small positive lower-usage prior for private experts, not a
full target distribution over all experts. The actual expert distribution is
mainly determined by the task loss and sample-dependent gates.

For inter-level aggregation, AHRA uses the entropy-form regularizer
\begin{equation}
\mathcal{L}_{\mathrm{hier}}
=
\lambda_{\mathrm{hier}}
\sum_{\ell=1}^{L}
\rho_{\ell}\log(\rho_{\ell}+\epsilon).
\end{equation}
Since $\sum_{\ell}\rho_{\ell}\log\rho_{\ell}=-H(\rho)$, minimizing
$\mathcal{L}_{\mathrm{hier}}$ with $\lambda_{\mathrm{hier}}>0$ mildly encourages
distributed level usage and prevents premature collapse to a single level. It is
a level-diversity stabilizer rather than a sparsity or concentration penalty.

\subsection{Learning Objective}

The final prediction is produced by a task-specific head
$\hat{y}=f_{\mathrm{task}}(z_{\mathrm{final}})$. 
For classification, $f_{\mathrm{task}}$ is a two-layer MLP with dropout $0.1$
and outputs logits, for MOSI/MOSEI regression, it outputs a scalar sentiment
score. Raw logits are passed directly to BCEWithLogits or CE without an
explicit sigmoid or softmax. The overall objective is:
\begin{equation}
\mathcal{L}_{\mathrm{total}}
=
\mathcal{L}_{\mathrm{task}}
+
\mathcal{L}_{\mathrm{dec}}
+
\mathcal{L}_{\mathrm{exam}}
+
\mathcal{L}_{\mathrm{col}}
+
\mathcal{L}_{\mathrm{hier}}.
\end{equation}
The task loss is dataset-dependent:
\begin{equation}
\mathcal{L}_{\mathrm{task}}
=
\begin{cases}
\operatorname{BCEWithLogits}(\hat y,y),\\
\operatorname{CE}(\hat y,y),\\
\operatorname{MSE}(\hat y,y).
\end{cases}
\end{equation}
for multi-label classification, single-label classification and sentiment regression.
MM-IMDB uses BCEWithLogits for multi-label genre prediction. Food101,
MVSA-Single, and MIntRec use CE, MOSI/MOSEI use MSE and are evaluated by
MAE, Acc-7, and F1. All auxiliary losses are computed only over available
modalities and valid modality pairs. See Appendix~\ref{app:method_details} for details.
\begin{comment}
\subsection{Learning Objective}

The final prediction is produced by a task-specific head: $\hat{y}=f_{task}(z_{final})$. For classification, $f_{task}$ is a two-layer MLP with dropout 0.1 and outputs logits. For MOSI/MOSEI regression, it outputs a scalar sentiment score. No sigmoid or softmax is applied before BCEWithLogits or CE. 
The overall training objective is:
\begin{equation}
\mathcal{L}_{\mathrm{total}}
=
\mathcal{L}_{\mathrm{task}}
+
\mathcal{L}_{\mathrm{dec}}
+
\mathcal{L}_{\mathrm{exam}}
+
\mathcal{L}_{\mathrm{col}}
+
\mathcal{L}_{\mathrm{hier}}.
\end{equation}
The task loss is defined according to the dataset type:
\begin{equation}
\mathcal{L}_{\mathrm{task}}
=
\begin{cases}
\operatorname{BCEWithLogits}(\hat y,y),\\
\operatorname{CE}(\hat y,y),\\
\operatorname{MSE}(\hat y,y).
\end{cases}
\end{equation}
for multi-label classification, single-label classification and sentiment regression.
MM-IMDB uses sigmoid binary cross-entropy because each sample may have multiple
genre labels. Food101, MVSA-Single, and MIntRec use softmax cross-entropy. MOSI
and MOSEI use MSE for sentiment intensity regression and report MAE, Acc-7, and
F1. All auxiliary losses are computed over modalities available in the
corresponding dataset and over valid modality pairs.
    
\end{comment}
\section{Experiments}
\label{experiment}

\subsection{Datasets and Implementation Details}

We evaluate AHRA on six multimodal benchmarks, including four classification datasets 
(MM-IMDB~\cite{mm-imdb}, Food101~\cite{Food101}, MVSA-Single~\cite{MVSA-Single}, 
and MIntRec~\cite{MIntRec}) and two trimodal sentiment analysis datasets 
(CMU-MOSI~\cite{Cmu-mosi} and CMU-MOSEI~\cite{Cmu-mosei}). 
For the classification datasets, we follow EAU~\cite{EAU} and report accuracy, 
micro-F1, and macro-F1. For MOSI and MOSEI, we follow EMOE~\cite{EMOE} and report 
MAE, Acc-7, and F1. AHRA is implemented in PyTorch and trained on four NVIDIA A100 GPUs. We use a unified training configuration across datasets unless otherwise specified, 
with early stopping based on validation F1. Detailed dataset information is provided in Appendix~\ref{app:datasets}. We evaluate AHRA against representative baselines, detailed descriptions and citations are deferred to Appendix~\ref{APP:MODEL-ZOO}.

% Detailed information see in Appendix~\ref{app:datasets} and Appendix~\ref{app:hyperparams}.

\begin{table}
\centering
\caption{Performance comparison across three benchmarks (\%). AHRA results are averaged over five random seeds.}
\small
\setlength{\tabcolsep}{2.5pt} % 调整列间距，数字越小间距越窄
\renewcommand{\arraystretch}{1.0} % 调整行高

\begin{tabular}{l|cc|cc|cc}
\toprule
\multirow{2}{*}{\textbf{Methods}}
& \multicolumn{2}{c|}{\textbf{MM-IMDB}}
& \multicolumn{2}{c|}{\textbf{Food101}}
& \multicolumn{2}{c}{\textbf{MVSA-Single}} \\
& F1-mic & F1-mac
& Acc & F1
& Acc & F1 \\
\midrule
\midrule
ConCat   & 49.76 & 52.58 & 88.20 & 88.19 & 65.89 & 65.43 \\
LateFu  & 59.56 & 58.91 & 90.69 & 90.77 & 76.88 & 75.72 \\
GMU       & 63.00 & 54.10 & 90.38 & 84.86 & 75.31 & 70.82 \\
MGMU    & 66.30 & 61.60 & 91.05 & 90.33 & 76.03 & 74.99 \\
MMBT        & 66.80 & 61.80 & 94.10 & 91.28 & 78.50 & - \\
% MMIM        & - & - & - & - & 78.33 & 75.79 \\
% HCSCL    & 67.39 & 61.58 & - & - & 78.47 & 75.02 \\
TMC        & 66.80 & 61.80 & 89.86 & 89.80 & 76.06 & 74.55 \\
QMF         & - & - & 92.92 & 92.93 & 78.90 & 77.18 \\
EAU        & - & - & 93.20 & 93.18 & 79.15 & 78.36 \\
MLA        & - & - & 93.33 & - & \textbf{79.84} & - \\
BPMulT     & 68.90 & 58.70 & - & - & - & - \\
ARL    & - & - & 93.55 & 93.55 & - & - \\
\textbf{AHRA} & \textbf{69.18} & \textbf{63.24} & \textbf{95.06} & \textbf{94.21} & \textbf{79.84} & \textbf{78.59} \\
\bottomrule
\end{tabular}

\label{tab:main_results}
\end{table}

\begin{table}[t]
\small
\caption{Performance Comparison on the MIntRec (\%).}
  \centering
  \setlength{\tabcolsep}{3.5pt}
  \renewcommand{\arraystretch}{1.0}
  \begin{tabular}{l|cccc}
    \toprule
    \textbf{Methods} & \textbf{Acc.} & \textbf{F1} & \textbf{Pre.} & \textbf{Rec.} \\
    \midrule
    \midrule
    Concat   & 68.54 & 66.22 & 66.90 & 66.10 \\
    LateFusion & 69.21 & 67.01 & 67.88 & 67.13 \\
    MAG-BERT& 70.34 & 68.19 & 68.31 & 69.36 \\
    MuLT        & 72.58 & 69.36 & 70.73 & 69.47 \\
    MISA       & 72.36 & 70.57 & 71.24 & 70.41 \\
    DLF       & 72.42 & 70.68 & 71.26 & 70.39 \\
    TSDA     & 72.61 & 71.43 & 72.13 & 70.74 \\
    EMOE       & 72.58 & 71.47 & 72.08 & 70.86 \\
    \textbf{AHRA}  & \textbf{73.02} & \textbf{72.16} & \textbf{72.97} & \textbf{71.36} \\
    \bottomrule
  \end{tabular}
  
   \label{tab:MIR}
\end{table}

\begin{table}[t]
\small
\caption{Performance Comparison on MOSI and MOSEI. $\downarrow$ indicates that lower values are better.}
\centering
\setlength{\tabcolsep}{2.0pt} % 调整列间距，数字越小间距越窄
\renewcommand{\arraystretch}{1.0} % 调整行高
\begin{tabular}{l|ccc|ccc}
\toprule
\multirow{2}{*}{\textbf{Methods}} 
& \multicolumn{3}{c|}{\textbf{CMU-MOSI}} 
& \multicolumn{3}{c}{\textbf{CMU-MOSEI}} \\
 & MAE$\downarrow$ & Acc7(\%)& F1(\%)
 & MAE$\downarrow$& Acc7(\%)& F1(\%) \\
\midrule
\midrule
TFN & 0.947 & 31.9 & 77.95 & 0.572 & 51.6 & 78.96\\
MulT& 0.846 & 40.05 & 81.66 & 0.673 & 48.37 & 80.86 \\
PMR& 0.895 & 40.60 & 79.83 & 0.645 & 48.88 & 81.56 \\
MISA     & 0.788 & 41.27 & 82.43 & 0.594 & 51.43 & 82.13 \\
FDMER& 0.760 & 42.88 & 83.22 & 0.571 & 53.21 & 83.35 \\
DMD& 0.744 & 43.88 & 83.55 & 0.561 & 54.18 & 83.88 \\
MCIS& 0.756 & 43.58 & 83.85 & 0.557 & 53.85 & 84.34 \\
CGGM& 0.747 & 43.21 & 84.13 & 0.551 & 53.47 & 84.14 \\
ConFEDE & - &42.33&85.53 &-&54.51&85.84\\
DEVA& - & 46.33 & 84.45 & - & 52.28 & 83.75 \\
I2RL &-&47.06&86.21&-&54.72&86.46\\
DLF& 0.731 & 47.08 & 85.04 & 0.536 & 53.9 & 85.27 \\
EMOE& 0.710 & 47.7 & 85.4 & 0.536 & 54.1 & 85.3 \\
\textbf{AHRA}&\textbf{0.702} & \textbf{48.77} & \textbf{86.36} & \textbf{0.522} & \textbf{55.65} & \textbf{86.66} \\

\bottomrule
\end{tabular}
\label{tab:main}
\end{table}

\subsection{Comparison with the State-of-the-Art}
We report the mean over five random seeds, with standard deviations provided in Appendix~\ref{app:Statistical}.

We first compare AHRA with existing methods on the four classification benchmarks. As shown in Table~\ref{tab:main_results} and Table~\ref{tab:MIR}, AHRA consistently matches or surpasses the strongest baselines. Specifically, on Food101, AHRA improves accuracy from 94.10
(MMBT~\cite{MMBT}) to 95.06, and F1 from 93.55
(ARL~\cite{ARL}) to 94.21. On MIntRec, AHRA reaches 73.02 accuracy and 72.16 F1, outperforming the strongest reported baselines of 72.61 accuracy (TSDA~\cite{TSDA}) and 71.47 F1 (EMOE~\cite{EMOE}). These results indicate that explicitly preserving modality-private information and routing it through private experts is beneficial for both image--text classification and trimodal intent recognition.

On the trimodal sentiment benchmarks CMU-MOSI and CMU-MOSEI, AHRA also achieves competitive or superior performance compared with recent multimodal models, as summarized in Table~\ref{tab:main}. On MOSI, AHRA reduces MAE compared with EMOE~\cite{EMOE}. On MOSEI, AHRA obtains the lowest MAE and the highest Acc-7 among the compared methods, while its F1 remains competitive with the strongest baseline. These results suggest that AHRA produces more informative multimodal representations for fine-grained sentiment prediction.

In summary, AHRA improves over prior methods on heterogeneous benchmarks by jointly factorizing features into shared and private streams and by using an adaptive expert hierarchy to select informative private tokens and semantic levels, thereby mitigating the dilution of modality specific cues inherent in single space fusion.

\begin{table}[t]
\caption{F1-macro on MM-IMDB under the train-test image/text availability settings reported by PEFT~\cite{PEFT}. Baseline values are quoted from the corresponding comparison table, while AHRA is evaluated under the same availability settings.}
\centering
\small
\setlength{\tabcolsep}{2.5pt} % 调整列距但仍尽量保持默认
\renewcommand{\arraystretch}{1.0} % 保持默认行高

\begin{tabular}{c|c|ccccc}
\toprule
\multirow{2}{*}{\shortstack{\textbf{Training}\\\textbf{Image/Text}}}
& \multirow{2}{*}{\shortstack{\textbf{Testing}\\\textbf{Image/Text}}}
& \multirow{2}{*}{\textbf{ViLT}}
& \multirow{2}{*}{\textbf{Lee}}
& \multirow{2}{*}{\textbf{Uni.}}
& \multirow{2}{*}{\textbf{PEFT}}
& \multirow{2}{*}{\textbf{AHRA}} \\
&&&&&& \\ % 这一空行让multirow的列自动撑开两行的高度
\midrule
\midrule
\multirow{4}{*}{100\%/30\%}
  & 100\%/30\% & 32.78 & 37.72 & 39.47 & 43.21 & \textbf{54.28} \\
  & 30\%/100\% & 26.58 & 21.68 & 42.12 & 54.67 & \textbf{62.93} \\
  & 65\%/65\%  & 30.55 & 30.80 & 40.99 & 49.09 & \textbf{58.15} \\
  & Average.   & 29.97 & 30.07 & 40.86 & 48.99 & \textbf{58.45} \\
\midrule
\multirow{4}{*}{30\%/100\%}
  & 100\%/30\% & 30.25 & 24.93 & 29.85 & 43.07 & \textbf{50.67} \\
  & 30\%/100\% & 37.97 & 47.10 & 54.37 & 56.03 & \textbf{63.37} \\
  & 65\%/65\%  & 34.45 & 36.76 & 37.61 & 49.81 & \textbf{53.96} \\
  & Average.   & 34.22 & 36.26 & 40.61 & 49.64 & \textbf{56.00} \\
\midrule
\multirow{4}{*}{65\%/65\%}
  & 100\%/30\% & 35.80 & 39.04 & 40.60 & 42.46 & \textbf{55.09} \\
  & 30\%/100\% & 36.65 & 42.68 & 53.19 & 55.26 & \textbf{63.43} \\
  & 65\%/65\%  & 36.66 & 41.33 & 47.34 & 49.24 & \textbf{58.65} \\
  & Average.   & 36.37 & 41.02 & 47.04 & 48.99 & \textbf{59.06} \\
\bottomrule
\end{tabular}

\label{tab:missing_mmimdb}
\end{table}

\begin{table}[t]
\small
\centering
\caption{Accuracy (\%) of different methods on MVSA-Single under various noise levels.}
\setlength{\tabcolsep}{1.5pt}
\renewcommand{\arraystretch}{1.0}

\begin{tabular}{l|c|cc|cc}
\toprule
\multirow{2}{*}{\textbf{Methods}}
& \textbf{Clean} 
& \multicolumn{2}{c|}{\textbf{Salt–Pepper}} 
& \multicolumn{2}{c}{\textbf{Gaussian}} \\
& $\epsilon$=0 & $\epsilon$=5 & $\epsilon$=10 & $\epsilon$=5 & $\epsilon$=10 \\
\midrule
\midrule
Concat  & 65.89 & 58.69 & 51.16 & 50.70 & 46.12 \\
Late Fusion & 76.88 & 67.88 & 55.43 & 63.46 & 55.16 \\
MMBT   & 78.50 & 74.07 & 51.26 & 71.99 & 55.34 \\
TMC   & 76.06 & 68.02 & 56.62 & 66.72 & 60.35 \\
QMF& 78.90 & 73.90 & 60.41 & 73.85 & 61.28 \\
EAU& 79.15 & 74.81 & 61.40 & 73.89 & 62.04 \\
\textbf{AHRA} & \textbf{79.84} & \textbf{76.43} & \textbf{62.20} & \textbf{73.98} & \textbf{63.37} \\
\bottomrule
\end{tabular}
\label{tab:noise_robustness}
\end{table}

% \subsection{Robustness to Noisy and Missing Modalities}
\subsection{Robustness Analysis}

We further evaluate AHRA under missing, imbalanced, and noisy modality conditions. On MM-IMDB, AHRA achieves the highest F1-macro across all image/text train--test availability ratios, including strongly mismatched distributions (Table~\ref{tab:missing_mmimdb}), showing its ability to adapt modality reliance when one stream is undersampled or partially absent. On MVSA-Single, AHRA also achieves the highest accuracy under all evaluated salt-and-pepper and Gaussian noise levels, maintaining the strongest absolute performance as corruption severity increases (Table~\ref{tab:noise_robustness}). These results indicate that shared--private separation, foreground-controlled private experts, and hierarchical co-fusion allow AHRA to suppress unreliable evidence and reallocate trust across modalities and semantic levels.

\begin{table}[t]
\caption{Ablation Studies of AHRA.}
\centering
\small
\setlength{\tabcolsep}{1.5pt}
\renewcommand{\arraystretch}{1.0}
\begin{tabular}{l|cc|cc}
\toprule
\multirow{2}{*}{Setting}
    & \multicolumn{2}{c|}{CMU-MOSI}
    & \multicolumn{2}{c}{Food101} \\
    & MAE $\downarrow$ & Acc-7(\%) & Acc(\%) & F1(\%) \\
\midrule
\midrule
\textbf{AHRA (Ours)} 
& \textbf{0.702} & \textbf{48.77} & \textbf{95.06} & \textbf{94.21} \\
\midrule
\multicolumn{5}{c}{\textit{\textbf{Importance of AHRA Components}}} \\
\textit{w/o} AEA   
& 0.724 & 47.10 & 94.02 & 93.30 \\
\textit{w/o} HMEC  
& 0.716 & 47.65 & 94.35 & 93.65 \\
\midrule
\multicolumn{5}{c}{\textit{\textbf{Importance of Modality}}} \\
\textit{w/o} Linguistic    
& 0.864 & 38.10 & 93.85 & 93.02 \\
\textit{w/o} Acoustic      
& 0.728 & 46.80 &  -   &  -   \\
\textit{w/o} Visual        
& 0.735 & 45.50 & 88.90 & 88.75 \\
\midrule
\multicolumn{5}{c}{\textit{\textbf{Importance of Regularization}}} \\
\textit{w/o} $\mathcal{L}_{\mathrm{dec}}$   
& 0.725 & 46.90 & 94.10 & 93.45 \\
\textit{w/o} $\mathcal{L}_{\mathrm{exam}}$  
& 0.716 & 47.40 & 94.32 & 93.70 \\
\textit{w/o} $\mathcal{L}_{\mathrm{col}}$   
& 0.708 & 48.15 & 94.72 & 93.98 \\
\textit{w/o} $\mathcal{L}_{\mathrm{hier}}$  
& 0.710 & 48.05 & 94.60 & 93.90 \\
\bottomrule
\end{tabular}
\label{tab:ablation-2}
\end{table}

\begin{table}[t]
\caption{Ablation on Fusion Mechanisms}
\centering
\small
\setlength{\tabcolsep}{1.5pt}
\renewcommand{\arraystretch}{1.0}
\begin{tabular}{l|cc|cc}
\toprule
\multirow{2}{*}{Setting}
    & \multicolumn{2}{c|}{CMU-MOSI}
    & \multicolumn{2}{c}{Food101} \\
    & MAE $\downarrow$ & Corr $\uparrow$ & Acc(\%) & F1(\%) \\
\midrule
\midrule
\textbf{AHRA (Ours)} 
& \textbf{0.702} & \textbf{0.806} & \textbf{95.06} & \textbf{94.21} \\
\midrule
\multicolumn{5}{c}{\textit{\textbf{Different Fusion Mechanisms}}} \\
Addition          & 0.746 & 0.762 & 94.05 & 93.28 \\
Multiplication    & 0.741 & 0.771 & 94.25 & 93.50 \\
CMAF & 0.729 & 0.785 & 94.55 & 93.82 \\
\bottomrule
\end{tabular}
\label{tab:fusion_mechanisms}
\end{table}

\subsection{Ablation Studies}

We conduct ablation studies on MOSI and Food101 to examine the contribution of each component in AHRA, with results reported in Table~\ref{tab:ablation-2} and Table~\ref{tab:fusion_mechanisms}.

\textbf{Effect of architectural components.}
Removing AEA consistently degrades performance on both datasets, showing that adaptive routing of private tokens through modality-specific experts is more effective than uniform private-feature processing. Removing HMEC also leads to performance drops, indicating that level-aware expert coordination is beneficial for jointly exploiting shared and private streams. These results confirm that AHRA preserves modality-private cues through the coupling of adaptive expert routing and hierarchical co-fusion.

\textbf{Effect of modalities.}
Dropping any modality weakens performance, while the full multimodal setting achieves the best results. On MOSI, removing language causes the largest degradation, consistent with its dominant role in sentiment prediction, whereas audio and vision provide complementary private cues. On Food101, removing vision leads to the most severe drop, reflecting the visually dominant nature of fine-grained food recognition. This demonstrates that AHRA can exploit dominant modalities while still benefiting from complementary modality-specific information.

\textbf{Effect of fusion mechanisms.}
Replacing HMEC with additive fusion, multiplicative fusion, or CMAF~\cite{FDMER} consistently underperforms the full model. This suggests that hierarchical expert-level fusion is more suitable than fixed operators or single-stage attention for determining which experts and semantic levels should contribute to prediction.

\textbf{Effect of regularization.}
Removing $\mathcal{L}_{\mathrm{dec}}$ causes the most evident degradation, verifying the importance of aligning shared streams while decorrelating private streams to avoid collapse into a single latent space. Removing $\mathcal{L}_{\mathrm{exam}}$ also hurts performance, indicating that sparse and confident foreground selection helps private experts focus on informative tokens. In contrast, dropping $\mathcal{L}_{\mathrm{col}}$ or $\mathcal{L}_{\mathrm{hier}}$ yields mild but consistent declines, suggesting that they mainly stabilize expert usage and level utilization rather than define the core representational capacity.

\begin{figure}
\centering
\includegraphics[width=0.95\linewidth]{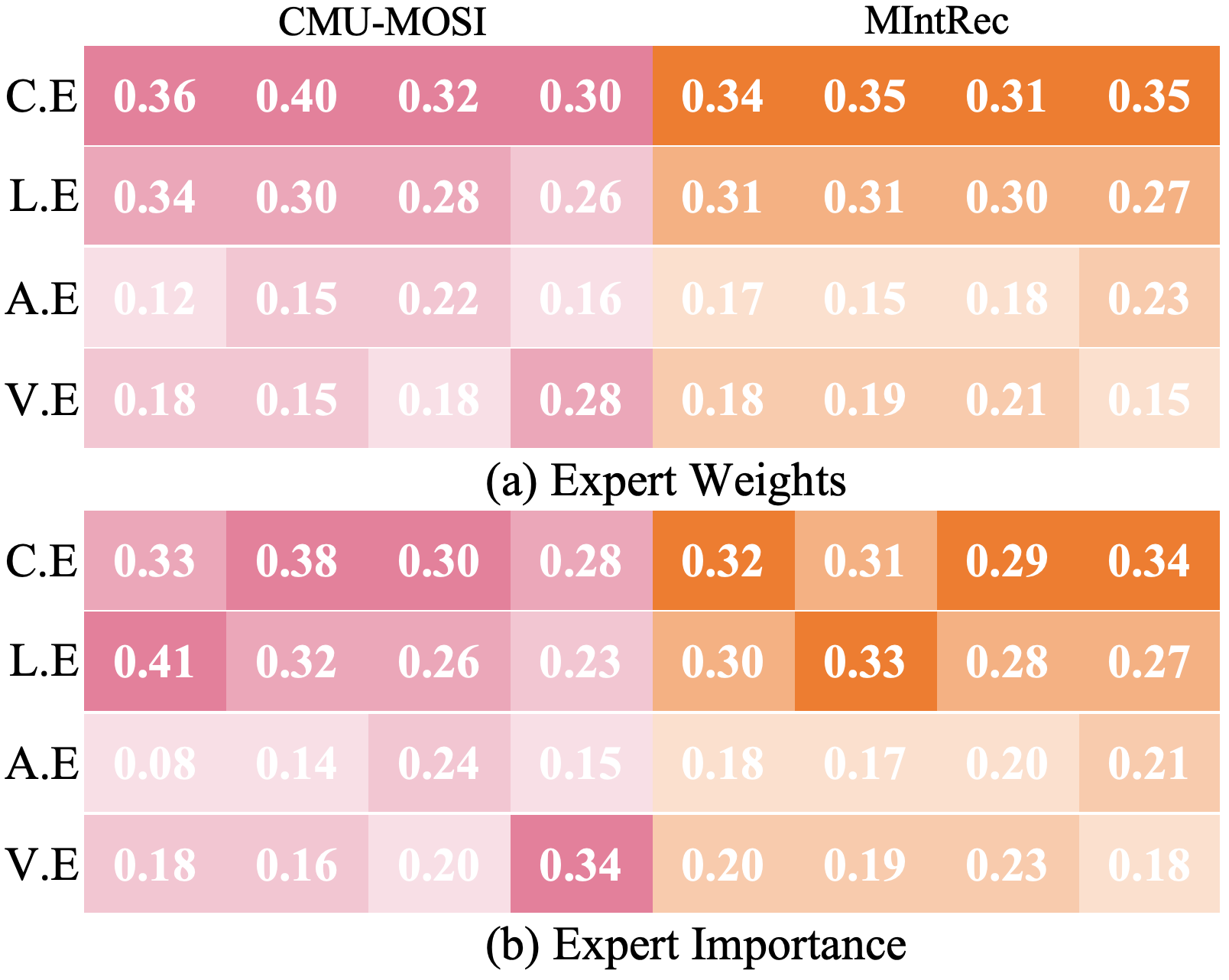}
\caption{Visualization of learned expert weights and importance for AHRA. CE and LE dominate with selective AE/VE activation on MOSI, while all experts contribute more evenly on MIntRec, showing that expert allocation adapts to both data and task.}
\label{fig:weight}
\end{figure}

\subsection{Visualization of Expert Weights and Importance.} 

To understand how AHRA allocates its experts, we visualize expert weights and expert importance on MOSI and MIntRec (Fig.~\ref{fig:weight}). The expert weights are the fusion coefficients produced by HMEC for the cross-modal expert (CE) and the modality-specific experts (LE, AE, VE), while the expert importance reflects their estimated contributions to the final prediction. On MOSI, CE and LE generally receive the largest weights and importance, whereas AE and VE are selectively amplified when prosodic or facial cues are salient. On MIntRec, the four experts are used in a more balanced manner, matching the setting where language, audio, and vision are all informative for intent recognition. Across both datasets, higher weights tend to coincide with higher importance, and different samples display distinct expert usage patterns. This indicates that AHRA exploits its expert bank in a data-dependent and task-aware way, rather than collapsing to a single dominant branch.

\begin{table}[!t]
\caption{Effect of the foreground exam budget on CMU-MOSI and Food101. ``FG ratio'' denotes the achieved average foreground rate under the target budget $\pi$.}
\centering
\small
\setlength{\tabcolsep}{3pt}
\renewcommand{\arraystretch}{1.0}
\begin{tabular}{c|c|cc|cc}
\toprule
Setting 
& FG ratio
& \multicolumn{2}{c|}{CMU-MOSI} 
& \multicolumn{2}{c}{Food101} \\
& (avg.) 
& MAE $\downarrow$ 
& Acc-7 (\%) 
& Acc (\%) 
& F1 (\%) \\
\midrule
\midrule
No exam 
& 100 
& 0.716 
& 47.40 
& 94.32 
& 93.70 \\
$\pi = 0.15$ 
& 17 
& 0.708 
& 48.20 
& 94.72 
& 93.98 \\
\textbf{$\pi = 0.30$ }
& \textbf{31} 
& \textbf{0.702} 
& \textbf{48.77} 
& \textbf{95.06} 
& \textbf{94.21} \\
$\pi = 0.45$ 
& 44 
& 0.707 
& 48.10 
& 94.90 
& 94.00 \\
$\pi = 0.70$ 
& 69 
& 0.712 
& 47.80 
& 94.65 
& 93.85 \\
\bottomrule
\end{tabular}
\label{tab:exam-budget}
\end{table}

\subsection{Analysis of the Foreground Exam Mechanism.} 

To clarify the role of the foreground exam, we vary a global foreground ratio $\pi \in \{0.15, 0.30, 0.45, 0.70\}$ in the exam loss while keeping all other components of AHRA fixed, and measure both the average fraction of private tokens emphasized by the soft foreground gate and performance on CMU-MOSI and Food101. We also include a variant without the exam, where all private tokens always enter the experts. As summarized in Table~\ref{tab:exam-budget}, the foreground budget controls the trade-off between selectivity and coverage: a very small budget ($\pi = 0.15$) activates too few private tokens and weakens modality-specific contributions, while a moderate budget ($\pi = 0.30$) yields the best results by filtering most background noise while retaining informative private tokens. Larger budgets ($\pi = 0.45$ and $\pi = 0.70$) gradually approach the no-exam variant, where many redundant or noisy private tokens are routed into the experts and performance declines. These results show that a moderate, sample-dependent foreground budget is crucial for selecting a compact yet informative subset of private tokens without either underusing private branches or overwhelming them with noise.

\begin{figure}
\centering
\includegraphics[width=1\linewidth]{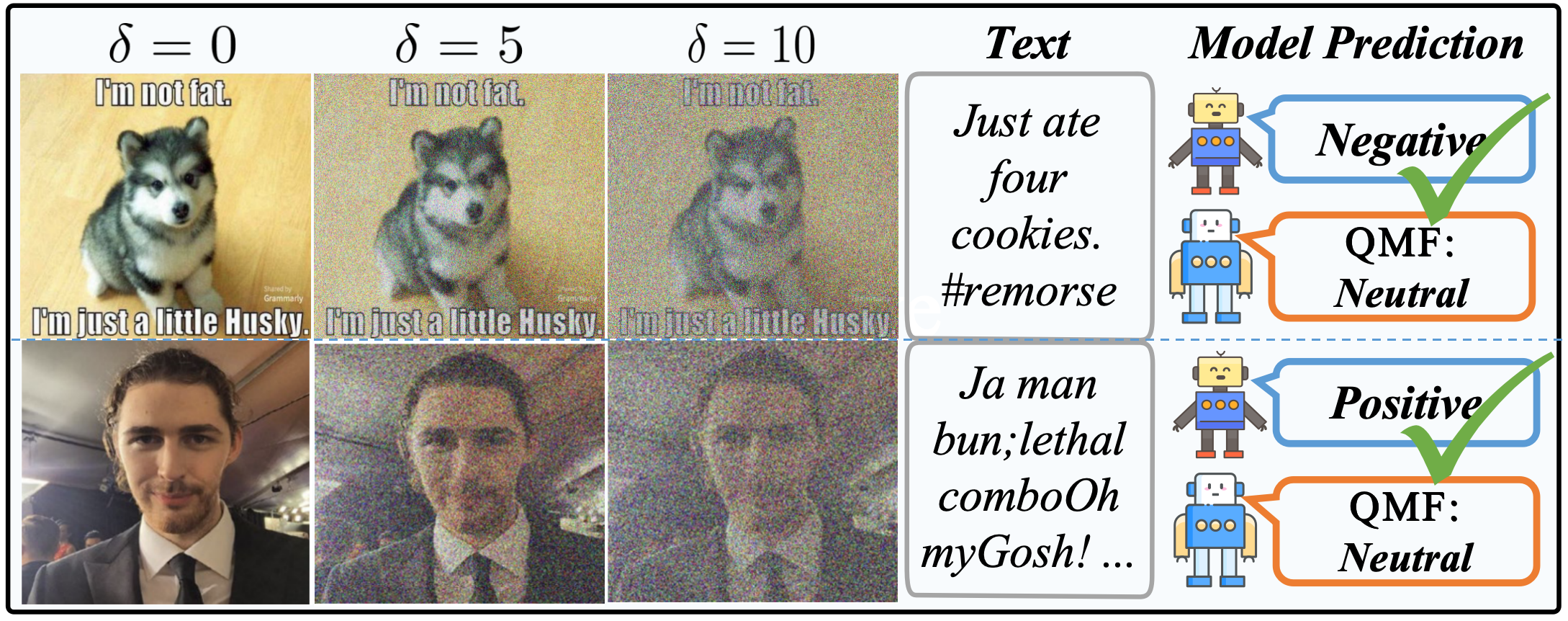}
\caption{Visualization of selected test cases from the MVSA-Single dataset. Our method consistently shows better robustness against noise in the three cases.}
\label{Qualitative Analysis}
\end{figure}

% \begin{table}[ht]
% \centering
% \caption{Computational efficiency analysis on CMU-MOSI. We report parameter counts and average training time per epoch.}
% \begin{tabular}{c|c c}
% \midrule
% \textbf{Model} & \textbf{Parameters} & \textbf{Time / Epoch} \\
% \midrule\midrule

% ConFEDE~\cite{confede}&246.98M & 40.12 s \\
% TFR-Net~\cite{TFR-Net} &- &- \\
% PRMF&117M & 18.62 s   \\
% LNLN&116M & 24.58 s  \\
% \textbf{AHRA (Ours)}&/ & / \\
% \midrule
% \end{tabular}
% \label{tab:Effciency}
% \end{table}

% \begin{figure}
% \centering
% \includegraphics[width=0.95\linewidth]{Attention Visualization.png}
% \caption{Qualitative multimodal recognition results of AHRA in a real-world sentiment analysis case.}
% \label{fig:Attention Visualization}
% \end{figure}

% \subsection{Attention Visualization}
% To illustrate the interactions between different modalities in AHRA, we present a speech visualization that integrates textual and visual features from the CMU-MOSEI dataset, as shown in Figure~\ref{fig:Attention Visualization}. The positive word "good" is highly correlated with visual features (e.g., the upturned corners of the mouth in the last image), while its correlation with features representing neutral facial expressions (e.g., the neutral facial expression in the first image) is weaker. This can be attributed to AHRA's ability to leverage the unique contributions of different modal features in the model to enhance and complement textual information.

\subsection{Qualitative Analysis}
\label{sec:qualitative_analysis}

% \textbf{Robustness Against Modality Noise.}
To examine the robustness of AHRA under noisy multimodal inputs, we visualize representative test cases with different noise levels in Fig.~\ref{Qualitative Analysis}. 
Compared with the baseline method QMF~\cite{QMF}, which is more sensitive to input perturbations and tends to produce incorrect predictions, AHRA maintains more stable and accurate sentiment predictions. This qualitative comparison indicates that the proposed AEA and HMEC modules help suppress modality-specific noise and preserve reliable cross-modal evidence, thereby improving prediction stability under noisy conditions. More qualitative results are provided in Appendix~\ref{app:more_results}.

\textbf{Complementary Analysis in the appendix further support AHRA}: t-SNE visualization shows more compact and class-separable embeddings (Appendix~\ref{app:t-sne}), sensitivity studies verify robustness to regularization coefficients (Appendix~\ref{app:Sensitivity}).

% and efficiency results confirm that AHRA achieves strong performance with controlled computational cost (Appendix~\ref{sec:Efficiency Analysis})

\section{Conclusion}
\label{section:Conclusion}

In this paper, we proposed the Adaptive Hierarchical Representation Alliance (AHRA) to mitigate the dilution of modality-private information in single-space multimodal fusion. AHRA factorizes each modality into shared and private streams across multiple semantic levels and couples an Adaptive Expert Alliance, a foreground exam, and a Hierarchical Multi-Expert Co-Fusion module to selectively enhance private cues while coordinating cross-modal evidence. Experiments on image–text classification, multimodal intent recognition, and trimodal sentiment analysis under noisy and missing modality conditions show that AHRA consistently outperforms strong baselines and ablations, confirming its effectiveness and robustness for multimodal recognition.

\section*{Limitations}
AHRA is evaluated on bimodal and trimodal benchmarks. Extending the framework to 
settings with many heterogeneous modalities, such as infrared, radar, depth, or 
physiological signals, may require additional modality selection or routing mechanisms 
to control computational cost and avoid expert over-fragmentation. In addition, AHRA 
does not explicitly address privacy risks, demographic bias, or dataset bias in 
multimodal data. These issues require further study before deployment in sensitive 
real-world applications. 

Our SGM analysis relies on CKA between layer representations and task-label kernels. CKA provides a useful diagnostic of representation-label similarity, but it does not establish causal evidence about which layer should be used for fusion and may miss nonlinear label dependencies not captured by the chosen kernels. Future work can combine CKA with probing classifiers or causal intervention analyses.

\section*{Ethics Statement}

This work studies multimodal representation learning on established academic benchmarks and does not introduce, collect, or release any new dataset. AHRA is evaluated on standard image--text classification, multimodal intent recognition, and trimodal sentiment analysis datasets, and all experiments are conducted for research purposes under their original benchmark settings. Since several benchmarks involve human-centered multimodal signals such as language, images, audio, or video, they may inherit annotation noise, demographic imbalance, privacy sensitivity, or social bias from the original data sources. Our method does not explicitly remove such dataset-level risks. 

AHRA aims to improve robustness by preserving shared and modality-private evidence across semantic levels and by reducing reliance on unreliable or corrupted modalities. However, robustness to noisy or missing modalities should not be interpreted as a guarantee of fairness, privacy protection, or deployment safety. In particular, sentiment and intent recognition models may be misused for intrusive profiling, surveillance, or consequential decision-making if applied without appropriate consent and oversight. Therefore, AHRA should be regarded as a research framework and should not be directly deployed in sensitive real-world applications without dataset-specific bias audits, privacy assessment, human supervision, and compliance with the usage policies of the corresponding data sources.

\bibliography{custom}
\appendix

\clearpage

\section{Related Works}
\label{app-section:Related work}

\subsection{Multimodal Learning}

Multimodal learning has been widely studied in tasks such as sentiment analysis, intent recognition, and vision language understanding. Recent studies have further extended multimodal reasoning to embodied and spatial settings, including vision-language-action planning and dynamic perception-memory integration~\cite{liao-1,liao-2}. Related diagnostic and alignment studies reveal persistent limitations in spatial world modeling and visual grounding, including reliance on non-visual priors and insufficient use of visual evidence~\cite{liao-3,liao-4,liao-6}. Early works such as TFN~\cite{TFN}, MMBT~\cite{MMBT}, and ViLT~\cite{ViLT} encode each modality separately and project the resulting features into a shared latent space, where a joint representation is learned for prediction. For video and audio–text modeling, MAG-BERT~\cite{MAG-BERT}, MulT~\cite{MuLT}, and other transformer-based models use cross-modal attention to capture temporal interactions between language, vision, and audio. Methods such as MISA~\cite{misa}, MCIS~\cite{MCIS}, CGGM~\cite{CGGM}, and DMD~\cite{dmd} explicitly model modality-invariant and modality-specific factors to obtain more structured multimodal representations for sentiment analysis. Most existing methods fuse all modalities in a single latent space, which biases representations toward shared factors and tends to wash out modality-private cues, especially under noisy or partially missing inputs. Despite their differences in architecture and training strategies, most of these methods ultimately concentrate both supervision and fusion in a single latent space. Strong cross-modal alignment or joint optimization tends to emphasize features that are predictive across modalities and for the target label, while subtle modality-private cues with weak cross-modal correlation are downweighted or compressed. As a result, the joint representation is often biased toward shared factors, and modality-specific information is easily diluted, especially under noisy or partially missing modalities.

\subsection{Multimodal Fusion Mechanism}

Many works design more flexible fusion mechanisms to handle modality heterogeneity. Simple concatenation and late fusion baselines~\cite{EAU} combine unimodal features at the feature or decision level, but ignore fine-grained cross-modal interactions. FDMER~\cite{FDMER} employs dynamic fusion with mutual enhancement and regularization, while MCIS~\cite{MCIS} and CGGM~\cite{CGGM} incorporate contrastive or graph-based modules to refine multimodal representations. EAU~\cite{EAU} explicitly models aleatoric uncertainty with stable unimodal augmentation and robust multimodal integration to obtain compact and noise-resistant joint representations. Robustness has also been examined at the representation level, where unified multimodal encoders exhibit modality-dependent vulnerability to adversarial
perturbations, motivating lightweight calibration of their shared representation spaces
\cite{liao-5}. EMOE~\cite{EMOE} integrates emotion-guided subspace interactions, DLF~\cite{DLF} adopts dual-level fusion blocks, and DEVA~\cite{DEVA} uses decoupled visual–acoustic streams with attentive fusion to enhance emotion recognition robustness. However, existing fusion mechanisms still fuse all modalities in a single joint embedding, which tends to suppress modality private cues.

\section{Methodological Details and Theoretical Rationale}
\label{app:method_details}

This appendix provides the details omitted from the main methodology because of
space constraints. The main paper defines the AHRA architecture and the training
objective; here we clarify the CKA diagnostic, the shared--private motivation, and
the exact interpretation of the correlation and gate regularizers.

\subsection{CKA Diagnostic for Semantic Granularity Mismatch}
\label{app:cka_sgm}

For each modality $m$ and semantic level $\ell$, we compute sample-level features
$\mathbf{h}^{(m)}_{\ell,i}=\mathrm{Pool}(H^{(m)}_{\ell,i})$ for all validation samples.
Let $X^{(m)}_{\ell}\in\mathbb{R}^{n\times d}$ stack these vectors. We use a centered
linear representation kernel
\begin{equation}
K^{(m)}_{\ell}=\Gamma X^{(m)}_{\ell}(X^{(m)}_{\ell})^{\top}\Gamma,
\qquad
\Gamma=I_n-\frac{1}{n}\mathbf{1}\mathbf{1}^{\top}.
\label{eq:app_rep_kernel}
\end{equation}
For single-label classification, the label matrix
$Y\in\mathbb{R}^{n\times C}$ is one-hot. For multi-label classification such as
MM-IMDB, $Y$ is multi-hot. In both cases, the label kernel is
\begin{equation}
K_y=\Gamma YY^\top\Gamma.
\end{equation}
For regression datasets such as CMU-MOSI and CMU-MOSEI, we use the scalar label
vector $y\in\mathbb{R}^{n}$ and set
\begin{equation}
K_y=\Gamma yy^\top\Gamma.
\end{equation}
The empirical CKA score is
\begin{equation}
\mathrm{CKA}\big(K^{(m)}_{\ell},K_y\big)
=\frac{\langle K^{(m)}_{\ell},K_y\rangle_F}
{\|K^{(m)}_{\ell}\|_F\|K_y\|_F}.
\label{eq:app_cka}
\end{equation}
This score is used only to diagnose the layer at which each modality becomes most
label-informative. It is not optimized during AHRA training and should not be read
as a proof of semantic granularity mismatch. The diagnostic supports the architectural
choice of retaining multiple semantic levels when different modalities peak at different
depths.

\subsection{Why Flat Alignment Can Suppress Private Factors}
\label{app:flat_alignment_proof}

We give a simple latent-factor argument for why single-space cross-modal alignment
can attenuate modality-private cues. Consider two modalities $m$ and $m'$ at one
semantic level, with zero-mean representations
\begin{align}
h^{(m)} &= A^{(m)}c+B^{(m)}u^{(m)}+\epsilon^{(m)},\label{eq:app_latent_m}\\
h^{(m')} &= A^{(m')}c+B^{(m')}u^{(m')}+\epsilon^{(m')},\label{eq:app_latent_mp}
\end{align}
where $c$ is a shared latent factor, $u^{(m)}$ and $u^{(m')}$ are modality-private
factors, and the noise terms are zero mean. Assume the shared, private, and noise
factors are mutually uncorrelated across the groups. A flat alignment objective with
linear projections $W_m$ and $W_{m'}$ is
\begin{equation}
\mathcal{J}=\mathbb{E}\left\|W_mh^{(m)}-W_{m'}h^{(m')}\right\|_2^2.
\label{eq:app_flat_align}
\end{equation}
Expanding Eq.~\eqref{eq:app_flat_align} under the uncorrelated-factor assumption gives
\begin{align}
\mathcal{J}
=&\left\|W_mA^{(m)}-W_{m'}A^{(m')}\right\|_{\Sigma_c}^{2}
\nonumber
\\
&+\left\|W_mB^{(m)}\right\|_{\Sigma_{u_m}}^{2}
+\left\|W_{m'}B^{(m')}\right\|_{\Sigma_{u_{m'}}}^{2}
\nonumber
\\
&+\left\|W_m\right\|_{\Sigma_{\epsilon_m}}^{2}
+\left\|W_{m'}\right\|_{\Sigma_{\epsilon_{m'}}}^{2},
\label{eq:app_flat_expanded}
\end{align}
where $\|Q\|_{\Sigma}^{2}=\mathrm{tr}(Q\Sigma Q^{\top})$. The shared component is
rewarded when $W_mA^{(m)}$ and $W_{m'}A^{(m')}$ match. By contrast, the private
components appear only as positive penalty terms because $u^{(m)}$ and $u^{(m')}$
are not predictable across modalities. Therefore, if the model relies only on flat
alignment and lacks a private path, minimizing alignment pressure can shrink the
private directions $W_mB^{(m)}$ and $W_{m'}B^{(m')}$. AHRA addresses this by keeping
private streams and routing them through modality-specific experts rather than forcing
all evidence into a single aligned space.

\subsection{Interpretation of the Decoupling Losses}
\label{app:decoupling_interpretation}

The shared alignment loss in the main paper uses dimension-wise batch standardization
followed by a cross-correlation target. This should be distinguished from full whitening.
Dimension-wise standardization ensures approximately zero mean and unit variance for
each coordinate in a mini-batch, but it does not remove all within-modality covariance.
Full whitening would require multiplying by an inverse covariance square root.

The finite-batch identity target should be interpreted as a regularizer rather
than an exactly attainable constraint. Let
$Z_m,Z_{m'}\in\mathbb{R}^{B\times d_{\mathrm{sh}}}$ be the standardized shared
representations in a mini-batch. The empirical matrix
$C=Z_m^\top Z_{m'}/B$ satisfies
\begin{equation}
\operatorname{rank}(C)
\le
\min(\operatorname{rank}(Z_m),\operatorname{rank}(Z_{m'}))
\le B.
\end{equation}
With centered standardized features, the effective rank is at most $B-1$ in the
usual case. Therefore, when $d_{\mathrm{sh}}>B$, as in our default setting
$d_{\mathrm{sh}}=256$ and $B=32$, $C$ cannot equal the full-rank identity matrix
$I_{d_{\mathrm{sh}}}$ in a single mini-batch. The identity-shaped target is used
as a stochastic correlation-shaping surrogate that encourages coordinate
agreement and redundancy reduction over training.

The private loss penalizes linear cross-correlation between streams. Zero covariance
is equivalent to independence only under additional assumptions such as joint Gaussianity.
For general neural representations, $\mathcal{L}_{\mathrm{pr}}$ should be interpreted
as encouraging linear decorrelation and reducing duplicated linear factors, not as a
guarantee of full disentanglement.

\subsection{Interpretation of the Foreground Exam}
\label{app:foreground_exam}

The foreground exam produces sample-wise soft gates
$\beta^{(m)}_{\ell,i,t}$ over private tokens. The rate term in
$\mathcal{L}_{\mathrm{exam}}$ controls the expected number of private tokens
passed to the private expert, while the confidence term penalizes uncertain
values near $0.5$. These terms alone do not prove that selected tokens are
semantically foreground or task-relevant. Task relevance is learned through
gradients from $\mathcal{L}_{\mathrm{task}}$ and the subsequent expert/fusion
modules, while $\mathcal{L}_{\mathrm{exam}}$ controls sparsity and stability.

The shared stream used in the private expert is interpreted as contextual
conditioning. In the main text this is written as
$E^{(m)}_{\mathrm{pr},\ell}(R^{(m)}_{\ell,\mathrm{FG}};\,c^{(m)}_{\ell})$,
where $c^{(m)}_{\ell}=\operatorname{Ctx}(S^{(m)}_{\ell})$. This notation does
not introduce a new training objective; it clarifies that private-token
enhancement is anchored in $R^{(m)}_{\ell,\mathrm{FG}}$ while being calibrated
by same-level shared context.

\subsection{Interpretation of the Hierarchical Regularizer}
\label{app:hier_regularizer}

The main paper keeps the original inter-level regularizer
\begin{equation}
\mathcal{L}_{\mathrm{hier}}
=\lambda_{\mathrm{hier}}\sum_{\ell=1}^{L}\rho_{\ell}\log(\rho_{\ell}+\varepsilon).
\label{eq:app_hier}
\end{equation}
Ignoring the numerical constant $\varepsilon$, this is $-\lambda_{\mathrm{hier}}H(\rho)$,
where $H(\rho)$ is the entropy of the inter-level distribution. Since the training
objective is minimized and $\lambda_{\mathrm{hier}}>0$, this term encourages higher
entropy, i.e., more distributed level usage. It should therefore be described as a
level-diversity or anti-collapse stabilizer, not as a concentration or sparsity penalty.
The inter-level gate can still produce non-uniform level weights because the task loss
and the gate network remain sample-dependent.

\subsection{Training Procedure}
\label{app:training_procedure}

For each mini-batch, AHRA is optimized as follows.
\begin{enumerate}
    \item Extract multi-level unimodal features $\{H^{(m)}_{\ell}\}_{\ell=1}^{L}$ for all available modalities $m\in\mathcal{M}$.
    \item Project each level into shared and private streams using Eq.~\eqref{eq:shared_private}.
    \item Compute $\mathcal{L}_{\mathrm{sh}}$ and $\mathcal{L}_{\mathrm{pr}}$ from mini-batch cross-correlations.
    \item Apply AEA: route shared streams through $E_{\mathrm{sh},\ell}$ and foreground-weighted private streams through $E^{(m)}_{\mathrm{pr},\ell}$.
    \item Apply HMEC: compute intra-level expert weights $\alpha_{e,\ell}$, level descriptors $\bar{z}_{\ell}$, inter-level weights $\rho_{\ell}$, and $z_{\mathrm{final}}$.
    \item Compute $\mathcal{L}_{\mathrm{task}}$, $\mathcal{L}_{\mathrm{exam}}$, $\mathcal{L}_{\mathrm{col}}$, and $\mathcal{L}_{\mathrm{hier}}$.
    \item Update all trainable parameters by minimizing $\mathcal{L}_{\mathrm{total}}$ with AdamW.
\end{enumerate}

\section{Dataset Details}
\label{app:datasets}

We evaluate AHRA on six multimodal benchmarks, including four classification datasets 
and two trimodal sentiment analysis datasets.

\textbf{Classification datasets.} The first group contains MM-IMDB~\cite{mm-imdb}, Food101~\cite{Food101}, 
MVSA-Single~\cite{MVSA-Single}, and MIntRec~\cite{MIntRec}. 
MM-IMDB consists of 25,959 movie samples, where each sample contains a movie poster 
and a synopsis for multi-label genre prediction. Food101 provides over 100,000 
image-text pairs from 101 food categories for fine-grained food classification. 
MVSA-Single contains 4,869 social media image--text pairs annotated with three-way 
sentiment labels. MIntRec includes 2,224 trimodal samples from 20 intent categories. Following EAU~\cite{EAU}, we report accuracy, micro-F1, and macro-F1 on these four datasets.

\textbf{Sentiment analysis datasets.} The second group contains CMU-MOSI~\cite{Cmu-mosi} and 
CMU-MOSEI~\cite{Cmu-mosei}. CMU-MOSI includes 2,199 monologue video clips, while CMU-MOSEI contains 22,856 video clips. Both datasets provide aligned language, acoustic, and visual features, together with sentiment scores in $[-3,3]$, ranging from strongly negative to strongly positive. Following EMOE~\cite{EMOE}, we report MAE, Acc-7, and F1 on these two datasets.

% \section{Implementation Hyperparameters}
% \label{app:hyperparams}

% To ensure reproducibility, we summarize the main hyperparameter settings used in our experiments in Table~\ref{tab:hyperparams}. Unless otherwise specified, the same configuration is adopted across all datasets.

% \begin{table}[t]
% \caption{Hyperparameter settings used in our experiments.}
% \label{tab:hyperparams}
% \centering
% \begin{tabular}{l c}
% \toprule
% Hyperparameter & Value \\
% \midrule
% Hierarchy levels $L$ & $3$ \\
% Internal dimensions $(d_{\mathrm{sh}}, d_{\mathrm{pr}})$ & $(256, 256)$ \\
% Foreground target ratio $\pi_{m,\ell}$ & $0.30$ \\
% Temperatures $(\tau_{\mathrm{intra}}, \tau_{\mathrm{inter}})$ & $(1.0, 1.0)$ \\
% Decoupling weights $(\lambda_{\mathrm{sh}}, \lambda_{\mathrm{pr}})$ & $(0.1, 0.1)$ \\
% Exam weights $(\lambda_{\mathrm{rate}}, \lambda_{\mathrm{conf}})$ & $(1.0, 0.1)$ \\
% Gate regularizers $(\lambda_{\mathrm{fair}}, \lambda_{\mathrm{hier}})$ & $(0.1, 0.1)$ \\
% Gate priors $(\eta_{\mathrm{L}}, \eta_{\mathrm{V}}, \eta_{\mathrm{A}})$ & $(0.1, 0.1, 0.1)$ \\
% Optimizer & AdamW \\
% Learning rate & $1\times10^{-5}$ \\
% Batch size & $32$ \\
% Max epochs & $35$ \\
% \bottomrule
% \end{tabular}
% \end{table}

\begin{figure*}
\centering
\includegraphics[width=0.97\linewidth]{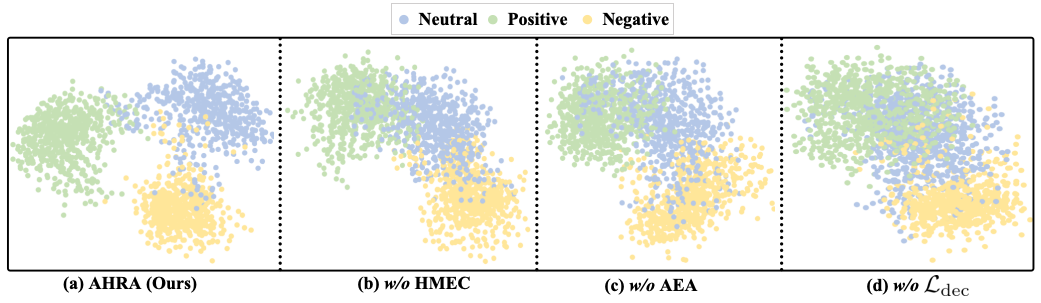}
\caption{t-SNE visualization of final multimodal embeddings on MVSA-Single under key AHRA designs. The full AHRA yields the most compact intra-class structures and the clearest inter-class separation. Removing HMEC weakens semantic-level aggregation, removing AEA reduces adaptive private-token enhancement, and removing $\mathcal{L}_{\mathrm{dec}}$ causes the strongest class overlap due to weakened shared--private structuring.}
\label{app-fig-tsne}
\end{figure*}

\begin{figure*}
\centering
\includegraphics[width=1\linewidth]{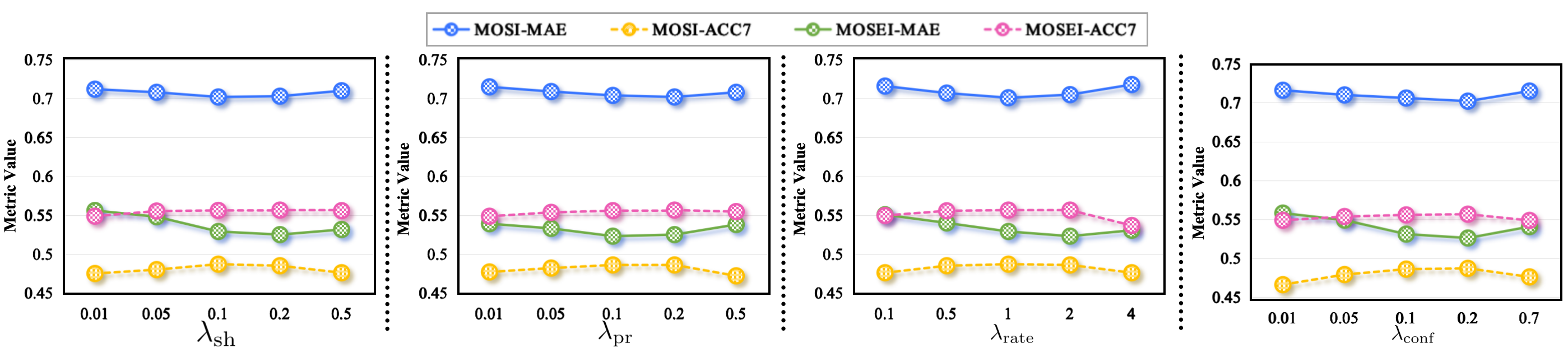}
\caption{Sensitivity of AHRA to the decoupling loss $\mathcal{L}_{\mathrm{dec}}$ and exam regularization $\mathcal{L}_{\mathrm{exam}}$ on CMU-MOSI and CMU-MOSEI.}
\label{fig:sen-1}
\end{figure*}

\begin{figure}
\centering
\includegraphics[width=1\linewidth]{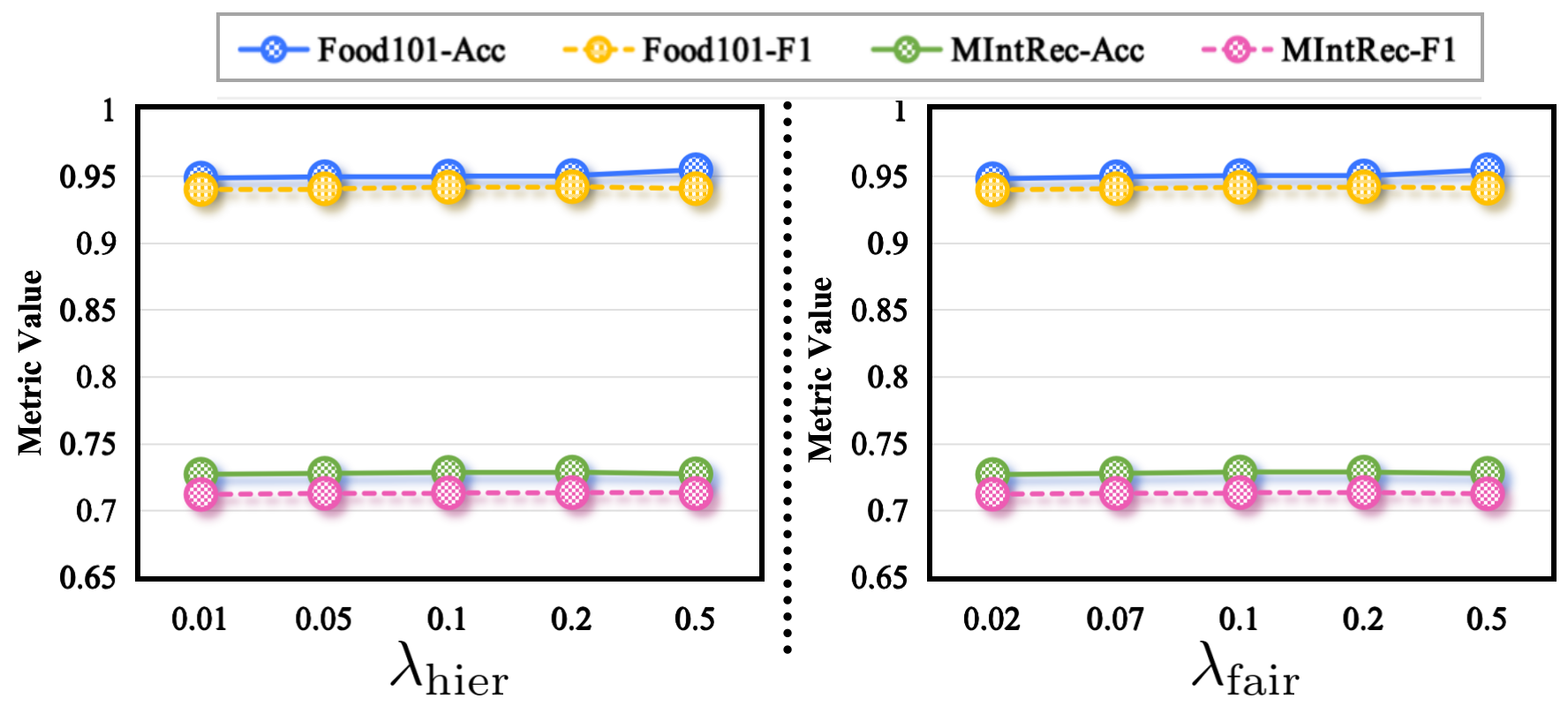}
\caption{Sensitivity of AHRA to the gating regularizers $\mathcal{L}_{\mathrm{col}}$ and $\mathcal{L}_{\mathrm{hier}}$ on the image--text and multimodal classification benchmarks Food101~\cite{Food101} and MIntRec~\cite{MIntRec}.}
\label{fig:sen-2}
\end{figure}

\section{Model Zoo}
\label{APP:MODEL-ZOO}
We compare AHRA with image-text classification baselines ConCat, LateFu, GMU, MGMU, MMBT, TMC, QMF, EAU, MLA, BPMulT, and ARL~\cite{EAU,GMU,MulT-GMU,MMBT,TMC,QMF,MLA,BPMulT,ARL}, multimodal intent baselines MAG-BERT, MulT, MISA, DLF, TSDA, and EMOE~\cite{MAG-BERT,MuLT,misa,DLF,TSDA,EMOE}, trimodal sentiment baselines TFN, PMR, FDMER, DMD, MCIS, CGGM, ConFEDE, DEVA, and I2RL~\cite{TFN,PMR,FDMER,dmd,MCIS,CGGM,confede,DEVA,I2RL}, and robustness baselines following PEFT~\cite{PEFT}. The tables therefore use compact method names without repeated citations.

\section{Statistical Reliability of AHRA}
\label{app:Statistical}

All AHRA experiments were repeated five times with different random seeds. As shown in Table~\ref{tab:ahra_full_stats}, we report mean and standard deviation for AHRA on all benchmarks, together with variance and 95\% confidence intervals to assess statistical stability. The statistical evidence confirms that AHRA exhibits high stability across modalities, dataset scales, and tasks, with consistently small variance and narrow confidence intervals. Combined with the ablation studies and robustness experiments, these results validate that the performance improvements of AHRA arise from principled architectural design rather than randomness.

\begin{table}[t]
\centering
\small
\caption{Statistical results of AHRA across all benchmarks. Mean and standard deviation are computed over 5 runs with different random seeds.}
\setlength{\tabcolsep}{1.5pt}
\renewcommand{\arraystretch}{1.0}
\begin{tabular}{l|cccc}
\toprule
\textbf{Dataset} & \textbf{Metric} & \textbf{Mean $\pm$ Std} & \textbf{Var.} & \textbf{95\% CI} \\
\midrule
\midrule
\multicolumn{5}{c}{\textbf{Image-text and Multimodal Classification (\%})} \\
\midrule
\midrule
MM
& F1-mic & 69.18 $\pm$ 0.33 & 0.11 & [68.77, 69.59] \\
-IMDB & F1-mac & 63.24 $\pm$ 0.41 & 0.17 & [62.73, 63.75] \\
\cmidrule{1-5}
Food101 
& Acc    & 95.06 $\pm$ 0.18 & 0.03 & [94.84, 95.28] \\
& F1     & 94.21 $\pm$ 0.21 & 0.04 & [93.95, 94.47] \\
\cmidrule{1-5}
MVSA
& Acc    & 79.84 $\pm$ 0.26 & 0.07 & [79.52, 80.16] \\
-Single & F1     & 78.59 $\pm$ 0.31 & 0.10 & [78.21, 78.97] \\
\midrule
\midrule
\multicolumn{5}{c}{\textbf{Multimodal Intent Recognition (MIntRec, \%)}} \\
\midrule
\midrule
MIntRec 
& Acc       & 73.02 $\pm$ 0.27 & 0.07 & [72.68, 73.36] \\
& F1        & 72.16 $\pm$ 0.30 & 0.09 & [71.79, 72.53] \\
& Precision & 72.97 $\pm$ 0.34 & 0.12 & [72.55, 73.39] \\
& Recall    & 71.36 $\pm$ 0.29 & 0.08 & [71.00, 71.72] \\
\midrule
\midrule
\multicolumn{5}{c}{\textbf{Trimodal Sentiment Analysis (MOSI / MOSEI)}} \\
\midrule
\midrule
CMU
& MAE   & 0.702 $\pm$ 0.006 & $3.6{\times}10^{-5}$ & [0.695, 0.709] \\
-MOSI & Acc-7 & 48.77 $\pm$ 0.37  & 0.14                & [48.31, 49.23] \\
& F1    & 86.36 $\pm$ 0.28  & 0.08                & [86.01, 86.71] \\
\cmidrule{1-5}
CMU
& MAE   & 0.522 $\pm$ 0.004 & $1.6{\times}10^{-5}$ & [0.517, 0.527] \\
-MOSEI & Acc-7 & 55.65 $\pm$ 0.25  & 0.06                 & [55.34, 55.96] \\
& F1    & 86.66 $\pm$ 0.19  & 0.04                 & [86.42, 86.90] \\
\bottomrule
\end{tabular}
\label{tab:ahra_full_stats}
\end{table}

\begin{figure}[htbp]
\centering
\includegraphics[width=0.95\linewidth]{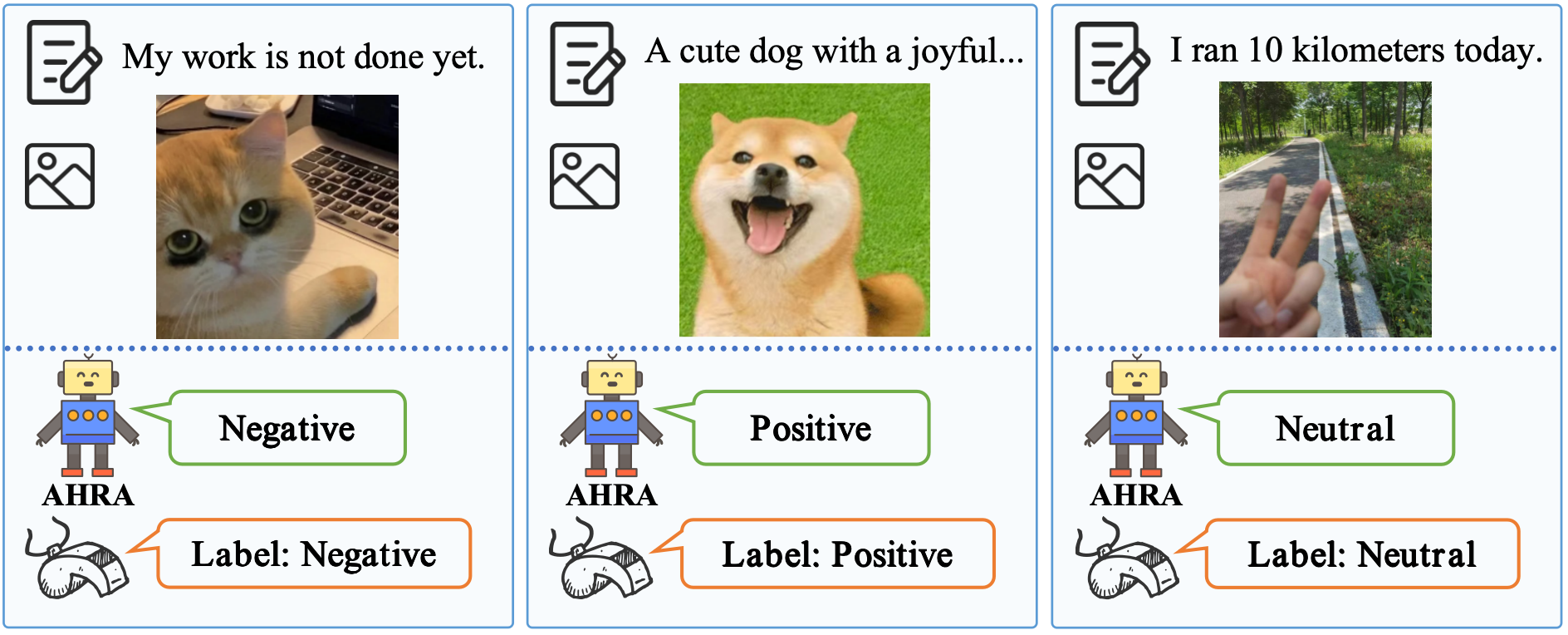}
\caption{Qualitative multimodal recognition results of AHRA in a real-world sentiment analysis case.}
\label{mvsa-3}
\end{figure}

\begin{figure}[htbp]
\centering
\includegraphics[width=0.95\linewidth]{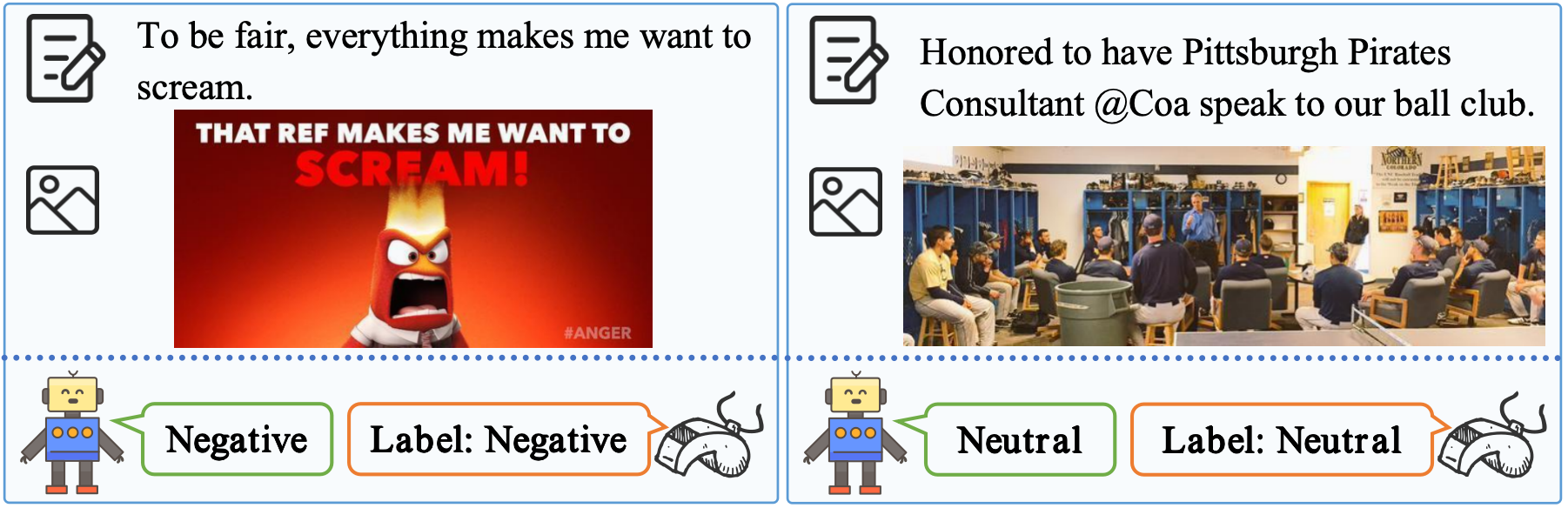}
\caption{Qualitative multimodal recognition results of AHRA on the sentiment analysis task.}
\label{mvsa-2}
\end{figure}

\section{Representation Visualization}
\label{app:t-sne}

To further examine the representation geometry learned by AHRA, we apply t-SNE~\cite{t-sne} to project the final multimodal embeddings on MVSA-Single into a two-dimensional space. As shown in Fig.~\ref{app-fig-tsne}, the full AHRA forms the most compact and class-separable sentiment clusters, indicating that the joint use of shared--private factorization, adaptive expert routing, and hierarchical co-fusion produces discriminative multimodal representations.

The ablated variants reveal how different designs contribute to this geometry. Removing HMEC leads to more diffuse decision regions and heavier boundary overlap, suggesting that intra-level expert coordination and inter-level semantic selection are important for organizing evidence across semantic levels. Removing AEA further weakens cluster compactness, which indicates that adaptive private-token enhancement helps preserve task-relevant modality-specific cues before final fusion. The variant without $\mathcal{L}_{\mathrm{dec}}$ exhibits the most severe class mixing, showing that shared alignment and private decorrelation are crucial for maintaining a structured shared--private representation space. These visual patterns are consistent with the quantitative ablation results and provide qualitative evidence that AHRA mitigates modality-private cue dilution while improving the discriminability of final multimodal embeddings.

\section{Sensitivity to Regularization Hyperparameters}
\label{app:Sensitivity}

We evaluate the sensitivity of AHRA to the main regularization hyperparameters that control shared–private decoupling, the foreground exam, and hierarchical gating. Instead of sweeping all coefficients on all datasets, we probe each group of terms on benchmarks where their effects are most directly reflected in the evaluation metrics.

We first study the decoupling loss $\mathcal{L}_{\mathrm{dec}}$ in Eq.~\eqref{eq:loss_dec} and the exam loss $\mathcal{L}_{\mathrm{exam}}$ in Eq.~\eqref{eq:loss_exam} on CMU-MOSI~\cite{Cmu-mosi} and CMU-MOSEI~\cite{Cmu-mosei}. These terms act on the representation level: $\lambda_{\mathrm{sh}}$ and $\lambda_{\mathrm{pr}}$ shape the alignment and separation of shared and private subspaces, while $\lambda_{\mathrm{rate}}$ and $\lambda_{\mathrm{conf}}$ control how strongly the foreground budget and gate confidence are enforced. Fine-grained sentiment regression on MOSI/MOSEI is particularly sensitive to such changes in token-level structure and modality-private cues. As shown in Fig.~\ref{fig:sen-1}, we therefore vary each of $\lambda_{\mathrm{sh}}, \lambda_{\mathrm{pr}}, \lambda_{\mathrm{rate}}, \lambda_{\mathrm{conf}}$ around the default setting, keeping the others fixed, and monitor MAE and Acc-7. Across a broad range of values, both datasets exhibit only mild fluctuations; performance degrades noticeably only when decoupling is almost removed or when the exam regularization becomes either too weak or too strong. This pattern is consistent with the design goal of AHRA: as long as a structured shared--private decomposition is preserved and a moderate fraction of private tokens is selected, the model maintains stable sentiment performance without aggressive tuning.

We then analyze the gating regularizers $\mathcal{L}_{\mathrm{col}}$ and $\mathcal{L}_{\mathrm{hier}}$ on Food101~\cite{Food101} and MIntRec~\cite{MIntRec}. These terms mainly stabilize private-expert usage and inter-level routing, rather than directly modifying token representations. Specifically, $\mathcal{L}_{\mathrm{col}}$ discourages systematic under-use of modality-specific experts, while $\mathcal{L}_{\mathrm{hier}}$ controls the strength of the level-diversity stabilizer. As shown in Fig.~\ref{fig:sen-2}, accuracy and F1 remain stable around the default setting, with degradation only at extreme values. Very small $\lambda_{\mathrm{fair}}$ can under-use private experts, whereas very large $\lambda_{\mathrm{fair}}$ forces nearly uniform expert usage and weakens adaptive routing. Similarly, removing or weakening $\mathcal{L}_{\mathrm{hier}}$ can make inter-level routing less stable, while an overly large $\lambda_{\mathrm{hier}}$ makes level usage too diffuse and  reduces the gate's ability to emphasize the most informative semantic scale for a given sample.

Overall, these studies indicate that AHRA is robust to the precise choice of regularization hyperparameters. The decoupling and exam coefficients can be tuned within a broad range without destabilizing fine-grained sentiment regression, and the gating regularizers admit similarly wide margins on diverse classification benchmarks. This robustness simplifies deployment in new multimodal settings, since reasonable default values already yield strong performance without extensive hyperparameter search.

\begin{figure}[htbp]
\centering
\includegraphics[width=1\linewidth]{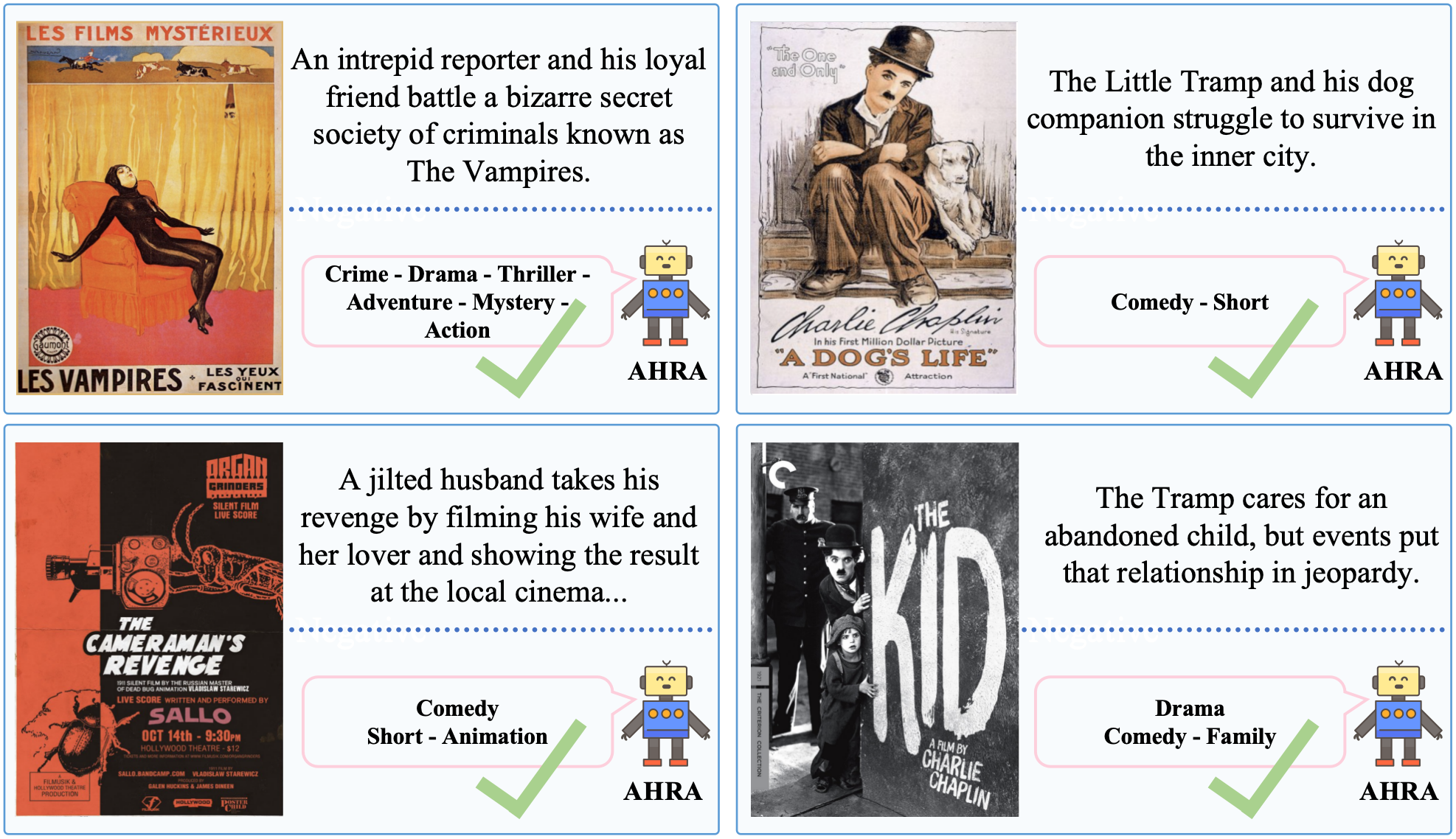}
\caption{Qualitative multimodal recognition results of AHRA on the movie category recognition task.}
\label{imdb-result}
\end{figure}

\section{Additional Qualitative Results}
\label{app:more_results}

\textbf{Versatility Across Semantic Domains.}
We further provide qualitative examples to examine the generalization capability of AHRA across different multimodal tasks, including sentiment analysis and movie genre recognition. 
As shown in Fig.~\ref{mvsa-3} and Fig.~\ref{mvsa-2}, AHRA can capture nuanced affective cues from complex real-world image-text scenarios, leading to reliable sentiment predictions. 
In addition, the results on the IMDb dataset in Fig.~\ref{imdb-result} show that AHRA can also support high-level semantic reasoning for multi-label movie genre prediction. 
These observations suggest that the adaptive expert collaboration and hierarchical fusion design enable AHRA to align and integrate heterogeneous multimodal evidence across diverse semantic domains.

\end{document}